\documentclass[12pt,preprint]{aastex}
\usepackage{hyperref}
\usepackage[all]{hypcap}
\hypersetup{
    bookmarks=true,         
    unicode=false,          
    pdftoolbar=true,        
    pdfmenubar=true,        
    pdffitwindow=false,     
    pdfstartview={FitH},    
    pdftitle={My title},    
    pdfauthor={Author},     
    pdfsubject={Subject},   
    pdfcreator={Creator},   
    pdfproducer={Producer}, 
    pdfkeywords={keyword1} {key2} {key3}, 
    pdfnewwindow=true,      
    colorlinks=true,        
    linkcolor=blue,         
    citecolor=blue,         
    filecolor=magenta,      
    urlcolor=cyan           
}

\slugcomment{}

\shorttitle{Sympathetic Eruption}
\shortauthors{Navin Chandra Joshi et al.}

\begin{document}

\title{Chain Reconnections observed in Sympathetic Eruptions}

\author{Navin Chandra Joshi\altaffilmark{1}, Brigitte Schmieder\altaffilmark{2}, Tetsuya Magara\altaffilmark{1,3}, Yang Guo\altaffilmark{4}, Guillaume Aulanier\altaffilmark{2}}
	\affil{$^1$ School of Space Research, Kyung Hee University, Yongin, Gyeonggi-Do, 446-701, Korea; navin@khu.ac.kr, njoshi98@gmail.com}
	\affil{$^2$ LESIA, Observatoire de Paris, UMR 8109 (CNRS), F-92195 Meudon Principal Cedex, France}
	\affil{$^3$ Department of Astronomy and Space Science, School of Space Research, Kyung Hee University, Yongin, Gyeonggi-Do, 446-701, Korea}
	\affil{$^4$ School of Astronomy and Space Science, Nanjing University, 210023 Nanjing, China}

\begin{abstract}

The nature of various plausible causal links between sympathetic events is still a controversial issue. In this work, we present multi--wavelength observations of sympathetic eruptions, associated flares and coronal mass ejections (CMEs) occurring on 2013 November 17 in two close--by active regions. Two filaments i.e., F1 and F2 are observed in between the active regions. Successive magnetic reconnections, caused by different reasons (flux cancellation, shear and expansion) have been identified during the whole event. The first reconnection occurred during the first eruption via flux cancellation between the sheared arcades overlying filament F2, creating a flux rope and leading to the first double ribbon solar flare. During this phase we observed the eruption of overlaying arcades and coronal loops, which leads to the first CME. The second reconnection is believed to occur between the expanding flux rope of F2 and the overlying arcades of the filament F1. We suggest that this reconnection destabilized the equilibrium of filament F1, which further facilitated its eruption. The third stage of reconnection occurred in the wake of the erupting filament F1 between the legs of overlying arcades. This may create a flux rope and the second double ribbon flare and a second CME. The fourth reconnection was between the expanding arcades of the erupting filament F1 and the nearby ambient field, which produced the bi-directional plasma flows towards both upward and downward. Observations and a nonlinear force-free field extrapolation confirm the possibility of reconnection and the causal link between the magnetic systems.

\end{abstract}


\section{Introduction}
\label{}

Solar filaments or prominences are characterized as cooler and denser plasma structures suspended in the hotter and more--tenuous corona. Filaments and prominences are considered as the same physical object but observed in the solar disk and off the solar limb, respectively. They always lie above the polarity inversion lines (PILs) at different locations on the Sun, for example, in the active regions, at the boundary of the active regions, or in the quiet regions, which are called respectively as active region filaments, intermediate filaments, and quiescent filaments \citep[][and references cited therein]{Labrosse10,Mackay10}.

Upward magnetic pressure, downward magnetic tension, gravity and the gas pressure gradient forces keep filaments in the equilibrium state in the chromosphere and in the corona \citep[ and references cited therein]{Forbes00,Schmieder13}. The filaments could undergo different kinds of eruptions such as full, partial, failed or asymmetric eruptions when the equilibrium breaks \citep{Gilbert07,Guo10b,Zhang15}. In a full eruption, the filament material and the associated magnetic field structure escape from the Sun and produce a coronal mass ejection \citep[CME;][]{Gilbert00,Schmieder15}. In a failed eruption, neither the filament material nor the associate magnetic field structure erupts \citep{Torok05,Liu08,Liu09,Guo10b,Joshi13a,Joshi14,Jiang14,navin14}. Various factors have been identified for the failed filament eruption such as a slow decrease of the overlying magnetic field \citep[e.g.,][]{Torok05}, a stronger overlying magnetic field \citep[e.g.,][]{Liu08} or an asymmetric background magnetic field \citep[e.g.,][]{Liu09}. More recently, \cite{Sun15} and \cite{Chen15} studied several confined flares and eruptions from AR 12192 during 2014 October 18 to 29 and found that the strong confinement of the overlying field may be the cause. In partial eruptions, some parts of the filament magnetic structure erupt while the other parts do not erupt, depending on the location of magnetic reconnection over or inside the filament \citep{Gilbert00,Gibson06,Tripathi09}. Asymmetric filament eruptions are the eruptions when one leg of the filament erupts while the other one remains anchored on the Sun \citep{Liu09,Joshi13a}.

Various factors have been observed to be responsible for the eruptions such as filament footpoint rotation \citep[e.g.,][]{Torok13}, flux emergence near the filament system \citep[e.g.,][]{Chen00}, large decrease of the photospheric magnetic field strength \citep[e.g.,][]{Schmieder08} or transfer of flux from one system to  another one \citep[e.g.,][]{Liu12}. Based on some observations and simulations, a number of theoretical models have been developed for the solar eruptions, such as the tether--cutting \citep{Moore01}, flux cancellation \citep{van89,Martens01}, magnetic breakout \citep{Antiochos99} and flux emergence models \citep{Chen00}. Ideal magnetohydrodynamic (MHD) instabilities has also been suggested as the cause for solar eruptions, i.e., kink and torus instabilities \citep[e.g.,][]{Torok05,Liu08}.

Successive solar eruptions and flares occurring within a short time with a physical link, either in one complex or in different active regions, are known as ``sympathetic events" \citep{Wang01,Moon02,Wang07,Jiang11}. Sympathetic eruptions have been observed and studied in the past years \citep{Zhukov07,Jiang09,Liu09,Jiang11,Schrijver11,Yang12,Shen12}. \cite{Wang01} studied two sympathetic M--class solar flares and associated CMEs that occurred on 2000 February 17 from two different active regions. \cite{Liu09} reported successive solar flares and associated two fast CMEs on 2005 September 13 from NOAA AR 10808 and discussed the causal relationship between these eruptions. \cite{Schrijver11} identified a series of flares and eruptions during 2010 August 1--2 and found that all these events are connected with a set of separatrices, separators, and quasi-separatrix layers. \cite{Jiang11} presented observations of three successive filament eruptions from different locations on 2003 November 19 and interpreted these sympathetic eruptions from coronal dimming investigations. \cite{Yang12} presented detailed observations of two filament eruptions on 2005 August 5 in a bipolar helmet streamer type configuration and interpreted these eruptions as sympathetic. \cite{Shen12} reported two sympathetic solar eruptions in a quadrupolar magnetic configuration occurred on 2011 May 12. In all these observational studies, they found some physical links between the two sympathetic events. Apart from observational works, there are a few simulations that have been performed to understand the sympathetic eruptions more clearly \citep[e.g.,][]{Torok11b,Lynch13}. \cite{Torok11b} simulated the sympathetic filament eruptions that occurred on 2010 August 1. \cite{Lynch13} presented a 2.5D MHD simulation of sympathetic magnetic breakout CMEs in a coronal pseudo--streamer type magnetic configuration. Among the various proposed interpretations and theories, some involve the removal of high--altitude coronal arcades through ideal or resistive processes \citep{Jiang09,Schrijver11,Torok11b,Yang12,Shen12}, while others involve sequential reconnections between adjacent flux systems \citep{Liu09,Chandra11,Liu14,Dalmasse15}.

Sympathetic eruptions and flares have been observed but the actual magnetic link and the physical mechanisms of these events are not fully clear. In this work, we present multiwavelength observations and interpretations of two successive eruptions from two different active regions that occurred on 2013 November 17. We discuss the multi-stages of the reconnections that occurred during these eruptions. We focus on the following main questions, (1) what are the reasons behind these sympathetic events? (2) how the reconnection occur between the arcades and/or the flux rope? (3) what are the role of these interactions/reconnections? We organize our paper as follows: Section~\ref{sect2} deals with the description of the observational data set used in the present paper. Magnetic structure and morphological evolution of the sympathetic eruptions, flares and associated CMEs are presented in Section~\ref{sect3}, where the nonlinear force-free field (NLFFF) extrapolation and the associated results are discussed in Section~\ref{sect38}. The main results and their discussions are presented in the last section (Section~\ref{sect4}).
\section{Observational Data Set}
\label{sect2}

We use the Global Oscillation Network Group (GONG) H$\alpha$ observations for the current study and collect data from the GONG archive (i.e., \url{http://halpha.nso.edu/archive.html}). It provides images at 6563 \AA~wavelength with a spatial resolution of $\rm 1\arcsec$ and a cadence of around 1 minute \citep{Har11}. The H$\alpha$ observations are used to study the filament eruption dynamics. For the multiwavelength analysis, we have also used the data observed by the Atmospheric Imaging Assembly \citep[AIA;][]{Lem12}, which provides full disk images in two Ultraviolet (UV), one white-light and seven Extreme Ultraviolet (EUV) wavelengths. AIA provides EUV images with a minimum cadence of 12 s and a pixel size of $\rm 0.6\arcsec$. Magnetic field data are observed by the Helioseismic and Magnetic Imager \citep[HMI;][]{Schou12}. HMI images have a spatial resolution of $\rm 0.5\arcsec$ with a minimum cadence of 45 s for the line-of-sight magnetic field and a minimum cadence of 720 s for the vector magnetic field. AIA and HMI are two of the instruments on board the {\it Solar Dynamics Observatory} (\textit{SDO}).

To study the kinematics of the eruptions in the limb view, we use the Sun Earth Connection Coronal and Heliospheric Investigation/Extreme Ultraviolet Imager (SECCHI/EUVI) data. It is an instrument on board the {\it Solar Terrestrial Relation Observatory} \citep[\textit{STEREO};][]{Wuelser04,Howard08}. It observes the Sun in four wavelengths, i.e., 304, 195, 284 and 171 \AA. \textit{STEREO-A}/EUVI images have a spatial resolution of $\rm 1.6\arcsec$. X-ray data have been taken from the {\it Reuven Ramaty High--Energy Solar Spectroscopic Imager} \citep[\textit{RHESSI};][]{Lin02} and the images have been reconstructed using the clean algorithm with an integration time of around 60 s. To study the CMEs associated with the eruptions, we have used the chronograph data observed by \textit{STEREO-A}/COR1 \citep{How08} and the Large Angle and Spectrometric Coronagraph \citep[LASCO;][]{Bur95} on board the {\it Solar and Heliospheric Observatory} (\textit{SOHO}).

\section{Multiwavelength Observations of the Sympathetic Events}
\label{sect3}

\subsection{Morphology and Magnetic Properties of the Filaments}
\label{sect31}

\begin{figure}
\vspace*{-3cm}
\centerline{
	\hspace*{0.0\textwidth}
	\includegraphics[width=2\textwidth,clip=]{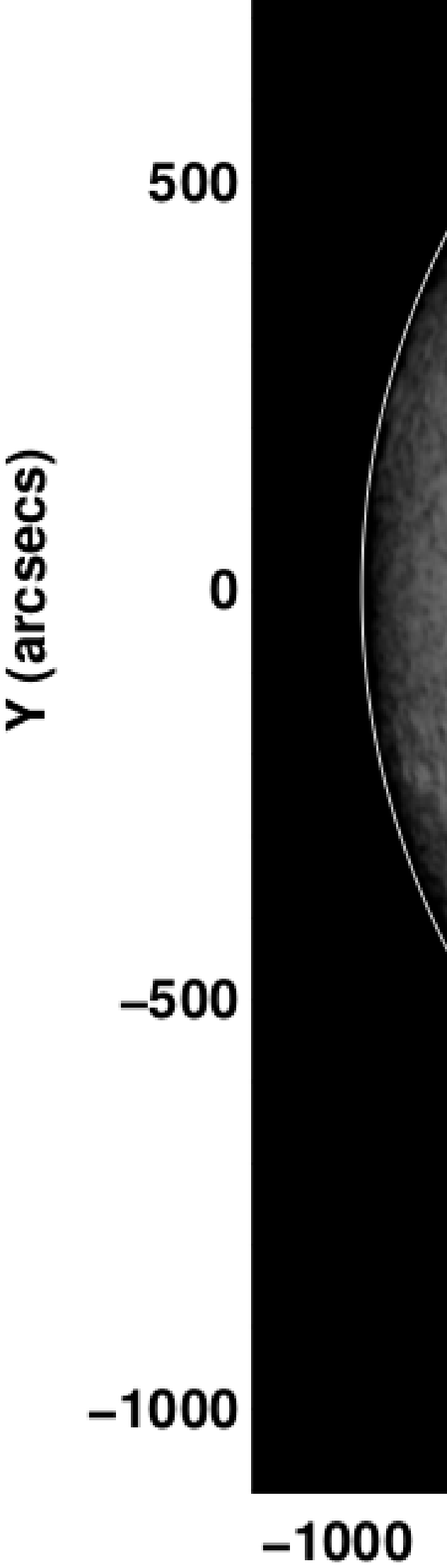}
	}
\vspace*{-2.5cm}
\caption{Full disk H$\alpha$ image of the Sun from Global Oscillation Network Group (GONG) on 2013 November 17 at $\sim$06:00:54 UT. White box shows the location of the two nearby filaments.}
\label{fig1}
\end{figure}

Two nearby filaments were observed on 2013 November 17 in the southwest part of the Sun (Figure~\ref{fig1}). The general structure of the filaments can be seen in Figure~\ref{fig2}. It displays GONG H$\alpha$, \textit{SDO}/AIA 304 and 171 \AA~zoomed images of the filaments at $\sim$06:00 UT (Figures~\ref{fig2}(a), \ref{fig2}(b) and \ref{fig2}(c) respectively). The northward and the adjacent southward filaments are represented by ``F1'' and ``F2'' respectively. They are at the border of two different active regions, i.e., NOAA 11893 and 11900 respectively. GONG H$\alpha$ image shows the location of the core dark filament material (Figure~\ref{fig2}(a)), while the extended filament channels can be seen in \textit{SDO}/AIA EUV images (Figures~\ref{fig2}(b) and~\ref{fig2}(c)). It can be seen that the dark filament material observed in H$\alpha$ lies in the middle of the extended filament channels. The filament channel could be the projection of the flux rope or cavity or it could be due to a different behavior of the transition region below the flux rope \citep{Aulanier02,Filippov15}. Different extrapolations of magnetic field lines in filament environment suggest that H$\alpha$ material of filaments is supported in the magnetic dips of a flux rope or sheared magnetic arcade \citep{Aulanier98,van04,Guo10}. The flux rope is overlaid by arcades in the corona and is anchored in the photosphere by barbs or footpoints created by parasitic polarities close to the PIL. The whole magnetic structure is called filament system.

The anchoring of the filament footpoints can be identified by comparing the H$\alpha$ and \textit{SDO}/AIA EUV observations with an \textit{SDO}/HMI magnetogram. The approximate locations of the footpoints are pointed out in Figure~\ref{fig2}(b) in which they are clearly visible. We have tracked the contours of the  approximate core filament region visible in H$\alpha$ of both filaments and over--plotted in the \textit{SDO}/HMI magnetogram (Figure~\ref{fig2}(d)). The magnetogram represents the line--of--sight component of the Sun's magnetic field. We have identified the footpoints of the two filaments. It is clear that the northern and southern foot points of both filaments are anchored in the positive and negative polarity regions in their respective active regions. The northern footpoints of both filaments lie near the center of the active regions, while the southern footpoints are anchored at the boundary of the respective active regions. It is also important to note that the filaments are  situated along the periphery of the active regions in the disk projection. The magnetic field between the two filaments is mostly consisting of positive polarity, while the other sides are mostly negative polarity.

The two filaments are located in the southern hemisphere, but they have different chirality. Filament F2 is dextral, or, equivalently the magnetic helicity is negative, because of the following reasons. First, the filament forms an inverse J shape as shown in Figure~\ref{fig2}. Secondly, according to the definition in \cite{Martin98a,Martin98b}, if one stands at the positive polarity, the axial magnetic field points to the right. Also, the filament barb is right bearing, although it does not have a one to one correlation with the filament chirality \citep{Guo10}. On the other hand, filament F1 is sinistral with a similar analysis as for filament F2. Therefore, the chirality of the two filaments is opposite. And the chirality of filament F2 is opposite to the dominant helicity sign of active regions in the southern hemisphere, which should be sinistral or positive in the term of magnetic helicity \citep[c.f.,][]{Martin98a,Martin98b,Pevtsov03,Lopez06}.

\subsection{Magnetic Field Changes in the Two Active Regions}
\label{sect32}

Figure~\ref{fig3} shows the radial component of the magnetic field at 00:10:11 UT on 2013 November 17. The magnetic field image has been corrected for the 180 degree ambiguity and the projection effect \citep{Hoeksema14}. We observe several magnetic field changes in both active regions before the start of the first eruption. All these changes can be seen very well in the \textit{SDO}/HMI radial component movie (movie attached with Figure~\ref{fig3}). Both active regions show several tiny flux emerging regions. These regions are shown by the northern and southern red ovals. These emerging regions show divergence motions as they expand.  In the northern active region, the  expansion towards the southwest disturbs the remnant network polarities such as the positive polarity in  box 2. We will see the importance of this polarity because magnetic field lines of both filament systems F1 and F2 are anchored in this polarity (see Section~\ref{sect38}). Moreover, in the northern active region, the following spot is in a decaying phase. It is surrounded by moving magnetic features, which form a moat region around it. This moat region is shown by a blue circle in Figure~\ref{fig3}. It is observed that many small bi--poles are canceling at the border of the moat region. One of the strong flux cancellation region is observed in box 1. This cancellation may trigger one reconnection (the third one probably).

\subsection{First Eruption and Associated Flare: First Stage Reconnection}
\label{sect33}

The first eruption stage was associated with the activation of filament F2 and the partial eruption of the overlying arcade. Figures~\ref{fig4} and~\ref{fig5} represent the sequence of selected H$\alpha$ and \textit{SDO}/AIA 304 \AA~images showing the overall chromospheric dynamics of the eruptions. Both the filaments were located near to each other and are represented by the letters ``F1'' and ``F2'' (Figures~\ref{fig4}(a) and~\ref{fig5}(a)). The southward confined motion of the cool plasma of filament F2 started at $\sim$06:45 UT toward its southern footpoint along the filament spine. The heating of the cool plasma (i.e., filament activation) started at $\sim$07:00 UT. The EUV brightness of the filament can be interpreted as the signature of heating (Figures~\ref{fig5}(a) and (b)). The confined southward movement of the cool and hot plasma continued until $\sim$07:15 UT (Figures~\ref{fig4}(a)--(c),~\ref{fig5}(a)--(c)). The partial disappearance of filament F2 during this confined motion has been observed in H$\alpha$ images (H$\alpha$ animation and Figure~\ref{fig4}(d)). This may also be due to the heating of the cool material of filament F2. The filament activation and heating are believed due to the reconnection underneath filament F2 between the legs of the overlying sheared arcades.

The reconnection dynamics can be clearly seen in \textit{SDO}/AIA 131 \AA~image sequence (Figures~\ref{fig6}(a)--(c)). The \textit{SDO}/AIA 131 \AA~channel provides information about the flaring reconnection region in the Sun. Activated filament F2 was visible at $\sim$07:00 UT (Figure~\ref{fig6}(a)). The approximate locations where the magnetic reconnection is supposed to occur between the legs of sheared arcades are represented by the white circle and oval in Figures~\ref{fig6}(a) and~\ref{fig6}(b) respectively. 
The reconnection continued and the formation of the flare loops has been observed at around $\sim$07:21 UT (Figure~\ref{fig6}(c)). These hot flare loops show continuous expanding motion with the eruption (animation associated with Figure~\ref{fig6}). After comparing the cooler and hotter AIA channels, we observed that these flare brightenings and loops only appeared in the hotter AIA channels i.e., 131 and 94 \AA. The formation of the flare ribbons on both sides of the reconnection region is clearly observed in H$\alpha$, \textit{SDO}/AIA 304 and 1600 \AA~images (Figures~\ref{fig4}(c)--(d),~\ref{fig5}(c)--(d) and~\ref{fig6}(d) respectively). These ribbons also show apparent separation motion away from each other (see H$\alpha$, \textit{SDO}/AIA 304 and 1600 \AA~animations associated with Figures~\ref{fig4},~\ref{fig5} and~\ref{fig6}, respectively). The northern ribbon moves toward north, while the southward ribbon moves toward the southwest. The flare loops connecting the flare ribbons can be evidenced by comparing \textit{SDO}/AIA 131 and 1600 \AA~images (Figures~\ref{fig6}(c) and~\ref{fig6}(d) respectively). The extension of the northern and southern ribbons of the first flare have a J shape and can be considered as the hooks of the flux rope containing the filament F2. The ribbons shown by the white ovals in Figure~\ref{fig6}(d) are the spreading footpoints of the growing flare loops.

In order to understand the temporal evolution of the EUV brightening during the flare events, we have estimated the EUV intensity profiles for the flaring region during 06:00 UT to 08:30 UT. These temporal profiles of \textit{SDO}/AIA 304, 131, 171 and 94 \AA~wavelength channels are represented in Figure~\ref{fig7}. For the comparison we have overplotted \textit{GOES} X-ray profiles in 1--8 \AA~and 0.5--4 \AA. Each profiles are normalized by different values for the better comparison of flare evolution. The region used for the estimation of these profiles are shown in Figure~\ref{fig5}(e) with a white box. The first flare phase can be considered from 07:00 UT to 07:35 UT. A gradual rise started at around 07:00 UT in all the EUV intensity profiles. The \textit{SDO}/AIA 304 and 171 \AA~profiles show a continues increase up to around 07:10--07:12 UT. These profiles show a gradual decrease up to 07:35 UT. \textit{SDO}/AIA 131 \AA~profile shows an initial gradual increase up to around 07:10 UT and then remain almost constant up to 07:35 UT. On the other hand, the 94 \AA~profile shows a consistent gradual increase with very slow rate up to 07:35 UT. A peak has also been observed in \textit{GOES} profiles, which appear to be consistent with the \textit{SDO}/AIA 304 and 171 \AA~profiles. The full disk observations at the same time reveal that this peak in the \textit{GOES} X-ray came both from the flare in this study as well as from another flaring activity occurred in another active region 11897. 

We do not see the eruption of the filament material during this phase in \textit{SDO}/AIA 304 \AA~and NSO/GONG H$\alpha$ observations. However, some evidence of the eruption from F2 and coronal loops can be found from \textit{SDO}/AIA 171 \AA \ and \textit{STEREO-A}/EUVI 195 \AA~images (Figures~\ref{fig8} and~\ref{fig9}). Figure~\ref{fig8} shows the \textit{SDO}/AIA 171 \AA~difference images, showing the eruption of filament F2 overlying loops between 07:09 UT and 07:15 UT in the disk view (see animation attached with Figure~\ref{fig12}). Erupting leading edge of the visible coronal loops is represented by dashed white curved lines. From the \textit{SDO}/AIA 171 \AA~disk observations, we also observe coronal dimming signatures just above filament F2 at 07:11 UT (Figure~\ref{fig8}(b)) as well as under the erupting loops at 07:14 UT (Figure~\ref{fig8}(c)) and 07:15 UT (Figure~\ref{fig8}(d)). The observed coronal dimmings show the observational evidence of eruptions from these regions. The locations of flare reconnection are also shown by the red circles in each panels. On the other hand, Figures~\ref{fig9}(a)--(e) show the \textit{STEREO-A}/EUVI 195 \AA~limb observations of the expansion and eruption of overlying coronal loops between 07:05:30 UT and 07:45:30 UT. The overlying loops are clearly visible in the limb view at 07:05:30 UT (Figure~\ref{fig9}(a)). The expansion and eruption of filament F2 overlying loops started at $\sim$07:05:30 UT and continued until 07:45:30 UT (animation associated with Figure~\ref{fig9}). The approximate leading edge of the erupting overlying loops is shown by the white arrow in each panel. The eruption of overlying coronal loops are very clear from the \textit{STEREO-A}/EUVI 195 \AA~movie. Figure~\ref{fig9}(f) shows the height--time profile of the erupting coronal loops using \textit{STEREO-A}/EUVI 195 \AA~images. A rough trajectory along which the height--time measurements are performed is shown in Figure~\ref{fig9}(d) with dashed white line. The error bars are the standard deviations estimated using three repeated measurements of the same point. The speed is estimated by the linear fit to these height-time data points. The estimated speed is around 160 $\rm km~s^{-1}$. We suggest that after the reconnection between the legs of the sheared arcades, some of the overlying arcades erupted. These arcade eruption results the expansion and eruption of the overlying coronal loops.

\subsection{Interaction between the Magnetic Field of the Two Filaments: Second Stage Reconnection}
\label{sect34}

Simultaneously with the first eruption, the ribbon separation of the first flare continued up to 07:32 UT, which has been discussed in detail in Section~\ref{sect33}. Looking at the 304, 94 and 131 movies, we can see the expansion of the northern ribbon of first flare, footprint of the hook of the F2 flux rope end. According to the motion and expansion of the hook of the F2 flux rope, it seems that the first reconnection is continuous even after the first flare and sucks the field lines between the two flux ropes, forcing some field lines to change of connectivity and reconnect far away. During the second stage interaction we observed several brightenings. At 07:32 UT, we observe the start of a first brightening under the filament F1 in H$\alpha$ and AIA EUV channels (Figures~\ref{fig4}(e),~\ref{fig5}(e),~\ref{fig10}(a) and (c)). This brightening corresponds to the slight bump in AIA 304, 131 and 171 \AA\ as well as the \textit{GOES} X-ray profiles during the intermediate time between the two flares (Figure~\ref{fig7}). This phase is represented by ``IF" in Figure~\ref{fig7}. This compact brightening is located at the region where the positive and negative polarities lie near to each other (i.e., corresponds to the box 1 in Figure~\ref{fig3}). It looks like the flux rope containing F1 is heated below and above F1, probably by a second reconnection. At 07:36 we see a second brightening in the south of the first one with a J shape in 304, 131 and 94 \AA\ channels (Figures~\ref{fig4}(e),~\ref{fig5}(e),~\ref{fig10}(d) and (f)). This could correspond to the flux rope hook of F1 and the elongated brightening could be considered as heated plasma along the flux rope. Thereafter the flux rope containing F1 is destabilized and heated. 

During the same time we note a brightening in the box (Figure~\ref{fig10}(c)) corresponding to the anchorage of the arcades overlying F1 and F2. The timing of the two brightenings suggest some heating all along the flux rope until its end in the place of the box. We also see some new brightenings above F1 like loops in 131 \AA\ at 07:37 and 07:41 UT (Figure~\ref{fig10}(d), (f), (g) and (i)). During the same time, the F1 flux rope (J ribbon) is heated in 304 \AA\ and starts to inflate from 07:41 UT. All these observational evidences suggest the second reconnection.

\subsection{Eruption of Filament F1 and Associated Flare: Third Stage Reconnection}
\label{sect35}

The activation of the northern filament F1 started at around $\sim$07:39 UT, a few minutes after the underneath heating. Thereafter, filament F1 started to rise between 07:42--07:43 UT (Figure~\ref{fig11}(a)). The eruption of filament F1 was associated with a small C4.9 class solar flare. The low lying reconnection between the legs of the overlying arcade field lines in the wake of filament F1 eruption may trigger this flare. This is the third stage of the magnetic reconnection. Thereafter, the filament F1 continued to erupt in the same direction as that of the first eruption (Figures~\ref{fig4}(f)--(g),~\ref{fig5}(f)--(g) and~\ref{fig11}(b)). We find the \textit{RHESSI} X-ray sources of 6--12 (red) and 12--25 (blue) keV over the flare brightening region at around 07:58 UT (Figure~\ref{fig11}(c)), which signifies the existence of the magnetic reconnection.

A careful inspection of the EUV intensity and \textit{GOES} X-ray profiles show that the third stage of the enhancement started at around 07:43 UT in all the EUV channels (Figure~\ref{fig7}). The \textit{GOES} X-ray flux enhancement also shows the third stage of the enhancement (red curves in Figure~\ref{fig7}). \textit{SDO}/AIA 304 and 171 \AA~profiles show early peaks at around 07:54 UT, while the \textit{SDO}/AIA 94 and 131 \AA~and \textit{GOES} X-ray profiles show the peaks of the flare between 08:00 UT and 08:05 UT. This third enhancement in the EUV profiles is due to the reconnection in the wake of the erupting filament F1 between the surrounding arcades. Moreover, the intensity profiles of \textit{SDO}/AIA 94 and 131 \AA\ and GOES X--ray channels also show another enhancement started at around 07:53:11 UT before reaching the peak. This provides evidence for another small reconnection stage during its eruption phase. NOAA record shows that the flare started at $\sim$07:37 UT, peaked at $\sim$08:01 UT and ended at $\sim$08:24 UT.

At about 07:58 UT, we observe that the erupting filament F1 broke into two parts, namely, the lower and the upper parts. These upper and lower parts are represented in Figures~\ref{fig4}(h),~\ref{fig5}(h) and~\ref{fig12}(a). Thereafter, we observe the eruption of the upper part only, while the lower part remains stable against eruption (animations associated with Figures~\ref{fig4} and~\ref{fig5}). Along with the eruption of the upper part, we observe the formation of two ribbons simultaneously (Figure~\ref{fig11}(d)). In the late decay phase at $\sim$08:30 UT, two set of post flare loops have also been observed in \textit{SDO}/AIA 171 \AA~channel (Figure~\ref{fig11}(e)). Right set of post flare loops (i.e., loops 1) are associated with the first flare, while the left set of post flare loops (i.e., loops 2) are associated with the second flare. The overall dynamics of both eruptions can be clearly seen in the H$\alpha$ and \textit{SDO}/AIA movies (see attached full length movies with Figures~\ref{fig4} and~\ref{fig5}).

\subsection{Reconnection between the Erupting Filament F1 and the Ambient Field: Fourth Stage Reconnection}
\label{sect36}

The erupting arcades of filament F1 underwent reconnection with the ambient field during $\sim$07:58 UT to $\sim$08:12 UT (Figure~\ref{fig12}). This can be interpreted as the fourth stage of the magnetic reconnection. Figure~\ref{fig12} represents the \textit{SDO}/AIA 171 \AA~running difference images during $\sim$08:01 to $\sim$08:12 UT. The eruptive upper and stable lower parts and the surrounding ambient field can be seen at 08:01 UT (Figure~\ref{fig12}(a)). The erupting upper part of filament F1 is reconnected with the nearby overlying ambient field lines (Figures~\ref{fig12}(b)--(d)). The approximate locations of the reconnection regions are marked by the red circles in each panel. As a result, the filament plasma appears to show downward and upward flow motions (animations associated with Figure~\ref{fig12}). The green arrows in Figures~\ref{fig12}(b)--(d) represent the directions of the upward and downward filament plasma motion. We also observe the formation of a new loop system joining the negative footpoint of filament F1 and the nearby positive polarity (Figures~\ref{fig4}(i),~\ref{fig5}(i) and~\ref{fig12}(b)--(d)).

\subsection{Kinematics of Filament F1 and Associated CMEs} 
\label{sect37}

Left and right panels of Figure~\ref{fig13} show the eruption of filament F1 using on disk observations by \textit{SDO}/AIA 304 \AA~and the limb observations by \textit{STEREO-A}/EUVI 304 \AA~respectively. The arrows in all the respective panels mark the similar leading edge of the erupting part of filament F1. Both the panels show similar morphological evolution of the filament F1 eruption. Figures~\ref{fig14}(a) and~\ref{fig14}(b) show the height--time profiles of the erupting part of filament F1 for on disk \textit{SDO}/AIA 304 \AA~as well as \textit{STEREO-A}/EUVI 304 \AA~ limb observations during 07:30 UT to 08:22 UT. The error bars in these plots are the standard deviations estimated by three repeated measurements of the same point. The rough trajectories along which the height--time measurements have been performed are represented in the Figures~\ref{fig13}(c) and~\ref{fig13}(f) with dashed white lines. In the height--time profile of \textit{SDO}/AIA 304 \AA,~we can identify two stages of the eruption of filament F1, namely, the first slow phase and the second fast phase. It is observed that initially filament F1 erupted with a slow speed during 07:41 UT to 08:02 UT (Figures~\ref{fig13}(a),~\ref{fig13}(b) and ~\ref{fig14}(a)). The estimated speed during this phase is around 145 $\rm km~s^{-1}$. It went into the fast phase eruption after 08:02 UT. The estimated higher speed is around 337 $\rm km~s^{-1}$ during 08:02 UT to 08:22 UT (Figures~\ref{fig14}(a)). The speed increase is due to the reconnection between the erupting filament F1 with the ambient magnetic field. The reconnection between the erupting filament F1 with the ambient field started at around 08:02 UT and continued up to around 08:15 UT, which is consistent with the fast rising phase of filament F1. Figure~\ref{fig14}(b) shows the height-time relationship using the \textit{STEREO-A}/EUVI 304 \AA~observations during 07:40 to 08:18 UT. The estimated speed is $\sim$247 $\rm km~s^{-1}$. Because of the low cadence, we can not see the two different phases in this panel. The speeds are calculated using the linear fit to the height-time data points during selected time intervals. The eruption of filament F1 seems to be asymmetric with the full eruptive northern and anchored southern legs (Figure~\ref{fig13}(f)).

Both the first and the second eruptions produced CMEs. The first of these eruptions originated from the vicinity of the filament we call F2, and produced the CME that we call CME1. During this first eruption, we observed disruption of filament F2 along with expulsion of loops above F2, but it is not clear whether F2 itself was ejected or if instead it just faded. Figures~\ref{fig8} and~\ref{fig12} show the erupting loops. The spreading ribbons in Figure~\ref{fig6}(d) are the footpoints of the arcade formed as the legs of these erupting loops reconnect, while the loops themselves continue to move outward to form part of CME1. The second eruption started with eruption of the filament we call F1, and produced the CME we call CME2. In this case, F1 clearly did erupt outward (e.g., Figures~\ref{fig5}(f) and~\ref{fig5}(g)). The flare ribbons shown in Figure~\ref{fig11}(d) are the footpoints of the arcade formed by this second eruption. Figure~\ref{fig11}(e) points out the post flare loops from both the first (post flare loops 1) and the second eruption (post flare loops 2).

Figure~\ref{fig15} represents the sequence of white-light chronograph images of the CMEs. Figures~\ref{fig15}(a)--(d) show the \textit{STEREO-A}/COR1 images from 07:25 UT to 08:25 UT, overplotted with the \textit{STEREO-A}/EUVI images. Figures~\ref{fig15}(e)--(h) represent the \textit{SOHO}/LASCO C2 and C3 images during 08:24 UT to 09:54 UT, overplotted with the \textit{SDO}/AIA images. All the panels show the complete sequence of the evolution of the CMEs. At $\sim$07:25 UT, we observe the dimming and the eruption of the overlying field in \textit{STEREO-A}/EUVI image (Figure~\ref{fig15}(a)). The eruption was observed in COR1 field of view at $\sim$07:45 in the form of CME1 (Figure~\ref{fig15}(b)). No core structure was associated with CME1. The erupting filament F1 was observed at $\sim$08:05 UT in the EUVI field of view while the leading edge of CME1 moved more outward in COR1 field of view (Figure~\ref{fig15}(c)). The erupting filament was seen for the first time in \textit{STEREO-A}/EUVI (Figure~\ref{fig15}(d)) as well as LASCO C2 (Figure~\ref{fig15}(e)) field of view at around 08:24--08:25 UT. The second CME was associated with a core structure, i.e., the erupting part of filament F1. At $\sim$08:48 UT we also observe multiple white and dark bands that show the signature of different leading edges and cavities (marked by black arrows in Figure~\ref{fig15}(f)). The similar pattern progressed simultaneously (Figures~\ref{fig15}(g) and~\ref{fig15}(h)), which suggests that both the CMEs produced by different eruptions merged and formed a compound CME structure at the later stages.

\subsection{Magnetic Field Modeling} \label{sect38}

In order to understand the overlying magnetic field configuration, we have carried out an NLFFF extrapolation using the optimization method \citep{Wiegelmann04,Wiegelmann06} over the active regions and presented in Figure~\ref{fig16}. Panels (a) and (b) show the approximate AIA on--disk and \textit{STEREO-A}/EUVI limb views respectively, while the bottom panel (c) represents the top view of the active regions with extrapolated field lines. Green and red lines represent the low lying magnetic field lines over the northern and southern interacting magnetic systems  respectively. Blue lines represent the open  field lines and the field lines not concerned by the present events. By comparing Figures~\ref{fig16}(c) and~\ref{fig2}(d), we note that the magnetic field lines in red are the overlying arcades of the filament F2, the magnetic field lines in green are the overlying arcade field of F1. Some connections between the overlying fields of F1 and F2 have also been observed. The white box shows the region of positive polarity between the active regions. As shown in the top view of the extrapolation result, some magnetic field lines overlying filament F2 extend to the region where some other magnetic field lines overlying F1 anchored. This signifies that magnetic interaction can occur between the two filaments. This connection is very important to explain the second stage of reconnection between the two filament magnetic systems. The second stage reconnection is supposed to occur between the expanding filament system (namely, the magnetic flux rope and surrounding field lines) of F2 and the overlying magnetic arcades of filament F1.

\section{Results and Discussions}
\label{sect4}
The present work deals with the observational analyses and interpretations of sympathetic eruptions, associated flares and CMEs using multiwavelength data from \textit{SDO}/AIA, \textit{GOES}, NSO/GONG, \textit{STEREO-A}/EUVI, \textit{STEREO-A}/COR1, \textit{RHESSI} and \textit{SOHO}/LASCO occurring in two active regions on 2013 November 17. We discuss various interactions and reconnections between filament magnetic field systems and between the filament magnetic field and the surrounding ambient magnetic field during the eruptions. The main results of the study are as follows:
\begin{enumerate}
\item Observations clearly show that the two successive eruptions and associated flares can be interpreted as the sympathetic events.
\item The magnetic reconnection can be divided into four stages during these sympathetic events. The first stage reconnection occurred as a flux cancellation between the legs of sheared arcades of filament F2. This might trigger a two ribbon solar flare, the eruption of the overlying coronal arcades and coronal loops.
\item The second stage reconnection could be between the field lines of the expanding flux rope F2 and the flux rope F1 as well as with its arcades. It activates F1 with heating signatures all along the flux rope and destabilizes it. 
\item The third stage reconnection occurred at the wake of the erupting filament F1 between the legs of overlaying arcades. This may trigger another two ribbon solar flare.
\item The fourth stage of reconnection occurred between the erupting arcades of the filament F1 with the nearby surrounding ambient field. We believe that the acceleration phase and the asymmetric eruption of filament F1 are the consequences of this coronal reconnection.
\end{enumerate}

\begin{figure}
\vspace*{-3cm}
\centerline{
	\hspace*{0.0\textwidth}
	\includegraphics[width=2\textwidth,clip=]{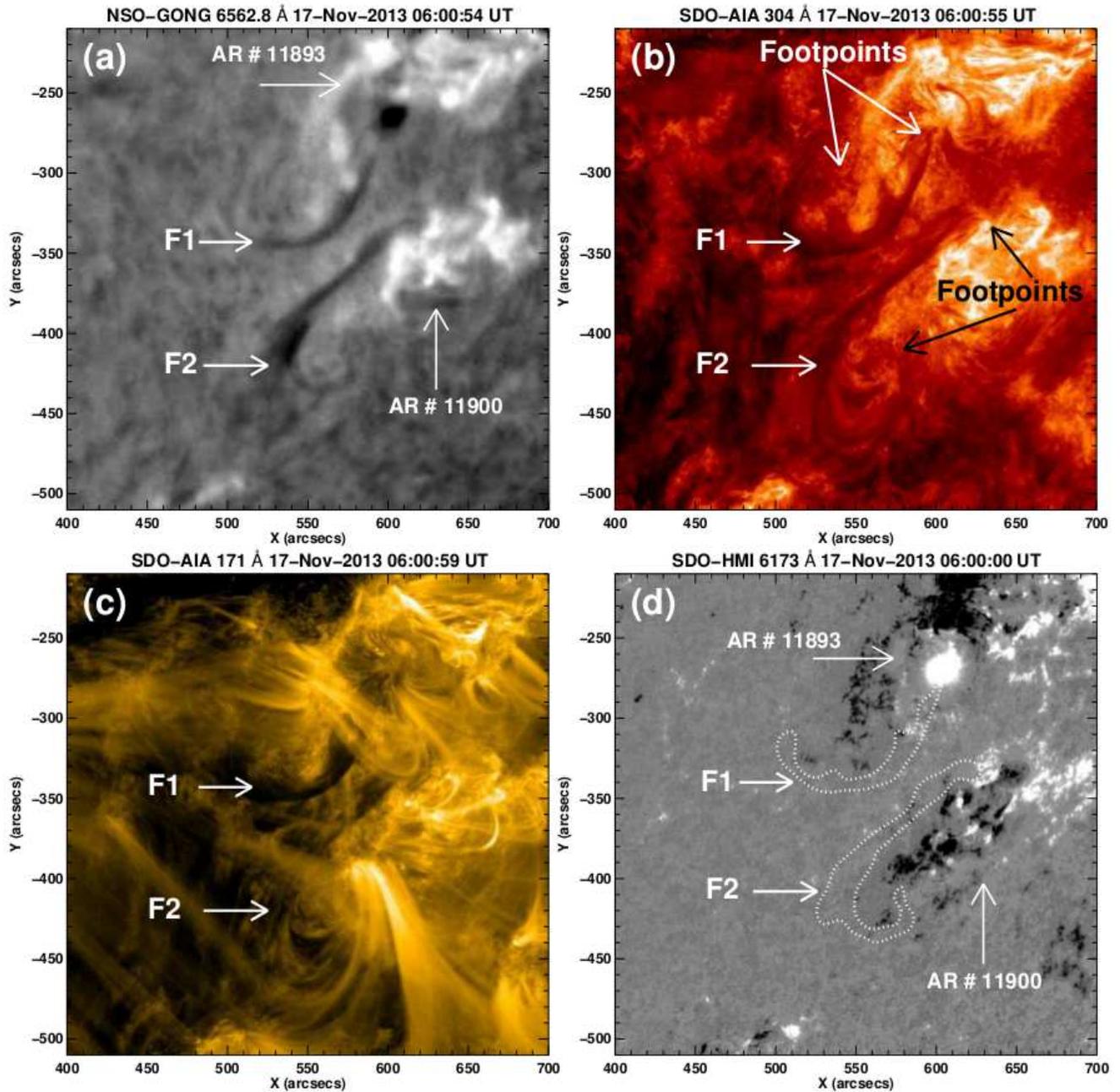}
	}
\vspace*{-1.4cm}
\caption{NSO/GONG H$\alpha$ (a), \textit{SDO}/AIA 304 \AA~(b) and \textit{SDO}/AIA 171 \AA~(c) images at $\sim$06:00 UT on 2013 November 17. Filaments are marked by the letter ``F''. The approximate location of footpoints of the two filaments are marked in panel (b). (d) The \textit{SDO}/HMI line--of--sight magnetogram overplotted by the tracked core filaments by the white dashed area. These core filament regions are extracted from the H$\alpha$ image.}
\label{fig2}
\end{figure}

\begin{figure}
\vspace*{-5cm}
\centerline{
	\hspace*{0.0\textwidth}
	\includegraphics[width=2\textwidth,clip=]{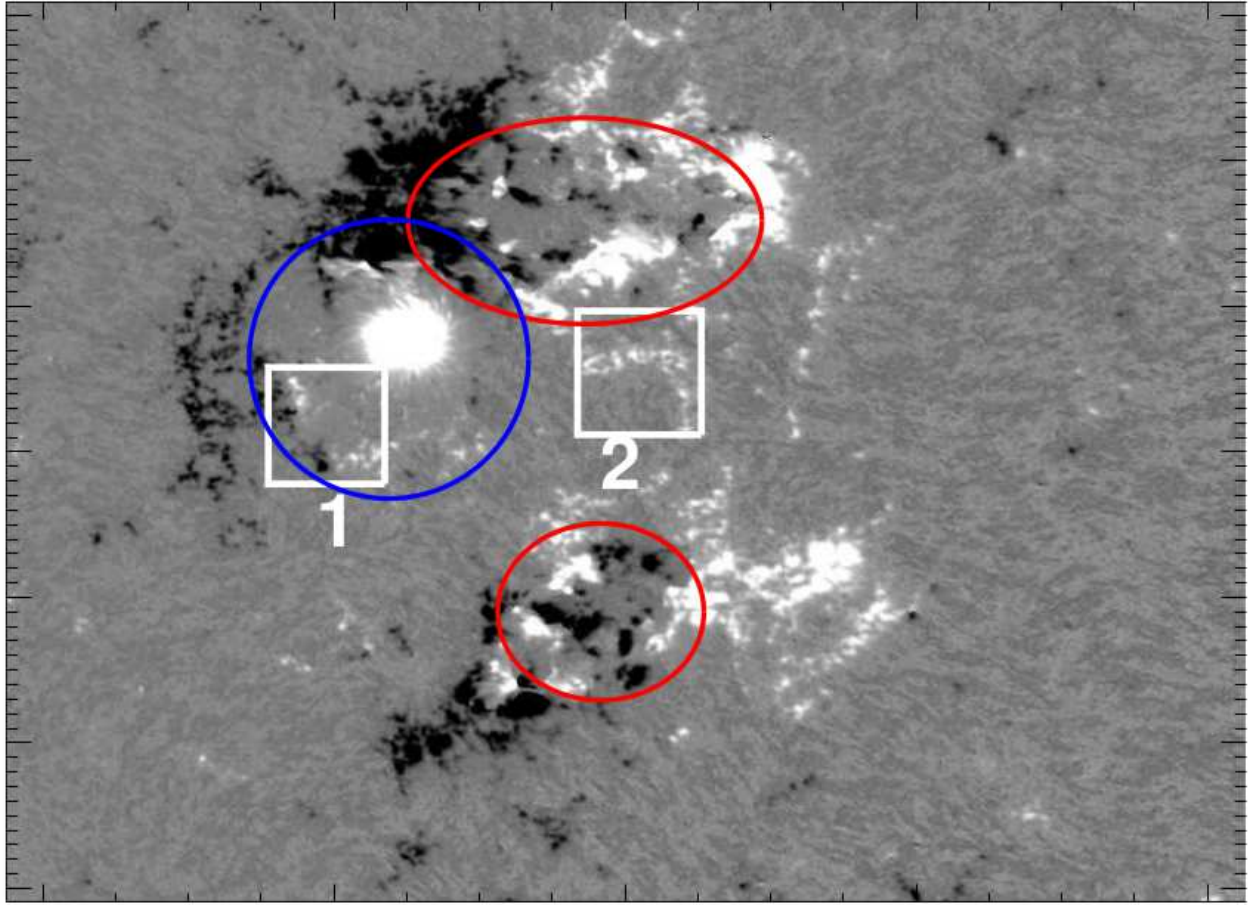}
	}
\vspace*{-3cm}
\caption{\textit{SDO}/HMI magnetogram (radial component) of both the active regions at 00:10:11 UT on 2013 November 17. The red ovals show the region of flux emergence regions in both  active regions. The blue oval shows the moat region around the positive polarity region where the flux cancellation is going on. Box 1 is the region where some of the flux cancellation has been observed near filament F1. Box 2 represents the area where the overlying field lines of both  filaments are anchored (see Figure~\ref{fig16}).}
\label{fig3}
\end{figure}

The present observed eruptions and flares can be interpret as two consecutive sympathetic eruptions and flares. For the sympathetic eruptions, it requires that they should occur in two different active regions and different PILs on the Sun with some physical connections \citep{Jiang09,Liu09,Jiang11,Torok11b,Yang12,Shen12,Lynch13}. In our case, we observed two eruptions, from two different nearby active regions as well as different PILs (Figure~\ref{fig2}). The NLFFF extrapolation shows the presence of overlying field lines of F1 and F2 anchored in the small positive region between them, showing some possible connection between them (see box 2 in Figures~\ref{fig3},~\ref{fig16} and~\ref{fig17}). The similar directions of both eruptions within a short time period of $\sim$40 minutes provides another evidence for sympathetic eruptions. We also observed two flares associated with these sympathetic eruptions, which can also be considered as the sympathetic flares. Two different set of flare loops have been observed close to each other with common anchoring at the central positive polarity region (Figure~\ref{fig11}(e)). Hence, the present observations not only show the sympathetic eruptions but also the sympathetic flares produced by them. Similar types of sympathetic flares associated with two eruptions have also been reported in a few earlier works \citep[e.g.,][]{Wang01,Wang07}.

Based on the multi-wavelength observations, we identify the four different stages of reconnections during the sympathetic eruptions (see schematic in Figure~\ref{fig17}). Figure~\ref{fig17}(a) represents the initial magnetic topology of the filaments and overlying ambient coronal fields. Two nearby filaments (i.e., F1 and F2) exist along with their overlying arcades as well as surrounding ambient fields (see Figure~\ref{fig2} also). The first flare reconnection seems to be consistent with the low--lying tether--cutting type reconnection between the sheared arcades leading to the formation of a flux rope \citep[c.f.,][]{Moore01}. The formation of the flare loops and their separation with time can be understood as resulting from reconnection between the legs of the surrounding field as the flux rope of that first eruption moves outward to form CME1. The growing flare loops and the spreading flare ribbons are the observational signatures of this first reconnection (see animations accompanying Figure~\ref{fig6}). Our observations show that the first eruption, which led to this first reconnection, began from near filament F2, but there is no clear ejection of F2. Apparently, F2 was located below the field that erupted in that first eruption. Similar kind of eruptions has been discussed in observational studies \citep[e.g.,][]{Pevtsov02,Tripathi09,Schmieder15}. Some of them have been interpreted with the partial eruption model of flux rope \citep{Gibson06,Tripathi09}.

The second stage of reconnection is the most difficult one to explain and it is not obvious where it happened. What we are sure is that there is an interaction between both systems (Figure~\ref{fig17}(d)) and the flux rope of F2 destabilizes the F1 system. Various observational signatures of second interaction and reconnection have been discussed in Section~\ref{sect34}. It has been observed that the removal of overlying field lines during the first eruption plays a crucial role to trigger a nearby eruption \citep[e.g.,][]{Jiang09,Jiang11,Yang12}. Some studies proposed that the external breakout type magnetic reconnection during the first eruption either in quadrupolar magnetic configuration \citep{Shen12} or in unipolar helmet streamer \citep{Torok11b,Yang12,Lynch13} magnetic configuration may be responsible for the other nearby successive eruption by removing the overlying flux. On the other hand the sucking of the field lines over F1 by the first reconnection is also a factor of diminishing the tension (green field lines over F2 in Figure~\ref{fig17}(c)). There was probably a driving reconnection pushing the two systems together. This kind of reconnection is only possible in 3D as shown MHD simulations for other event sequences \citep{Torok11b}.

The third stage of reconnection is believed to occur during the eruption of filament F1 and has some similarity with the first stage of reconnection occurring in the F2 system (Figures~\ref{fig11}(b)--(d) and~\ref{fig17}(e) and Section~\ref{sect35}). This reconnection may be consistent with the standard CSHKP flare model by filament or flux rope eruption \citep[c.f.,][]{Carmichael64,Sturrock66,Hirayama74,Koop76}. According to this, during the eruption of filament or flux rope the reconnection between the legs of surrounding arcades in the wake may trigger a solar flare.

The erupting arcades of filament F1 also reconnected with the near-by ambient magnetic flux (Figure~\ref{fig17}(f)) and formed another new loop system (Figure~\ref{fig17}(f)). We interpret this reconnection as the fourth stage of reconnection (refer to Section~\ref{sect36} for detail). This reconnection can be interpreted between the ambient open field lines with the filament F1 growing arcades in the corona. Similar kind of coronal reconnection senario has been discussed by \cite{Vrsnak03} to interpret the filament eruption on 2000 September 12. Recently, for the first time \cite{van14} found the evidence of reconnection in the corona between the field lines of two different active regions during CMEs. They also identified the bright coronal region of reconnection as well as the cool down flowing plasma. In our case, we also find observational signature of coronal reconnection via bi--directional flow of filament F1 plasma after their arcade reconnection with the ambient magnetic field (Figure~\ref{fig12} and associated animations). We also believe that this reconnection was responsible for the acceleration phase of the eruptive part of filament F1 (Section~\ref{sect37} and Figure~\ref{fig14}(a)). This result suggests that this kind of coronal magnetic field reconnection may play an important role in accelerating the filament speed during its eruption. This reconnection is also believed for the asymmetric eruption of filament F1. We observed that the northern leg of filament F1 erupted fully, while the southern leg remained anchoring on the Sun (Figure~\ref{fig13}). We believe that the interaction and reconnection of the erupting southern part of filament F1 with the ambient magnetic field caused the eruption asymmetry. This is a different mechanism compared with the whipping and zipping type model of asymmetric filament eruption given by \cite{Liu09}. In the end, three different types of loop system are formed, i.e., two post flare loops and one newly formed loop system (Figures~\ref{fig11}(e),~\ref{fig12}(c) and~\ref{fig17}(g)).

\begin{figure}
\vspace*{-4cm}
\centerline{
	\hspace*{0.0\textwidth}
	\includegraphics[width=2.2\textwidth,clip=]{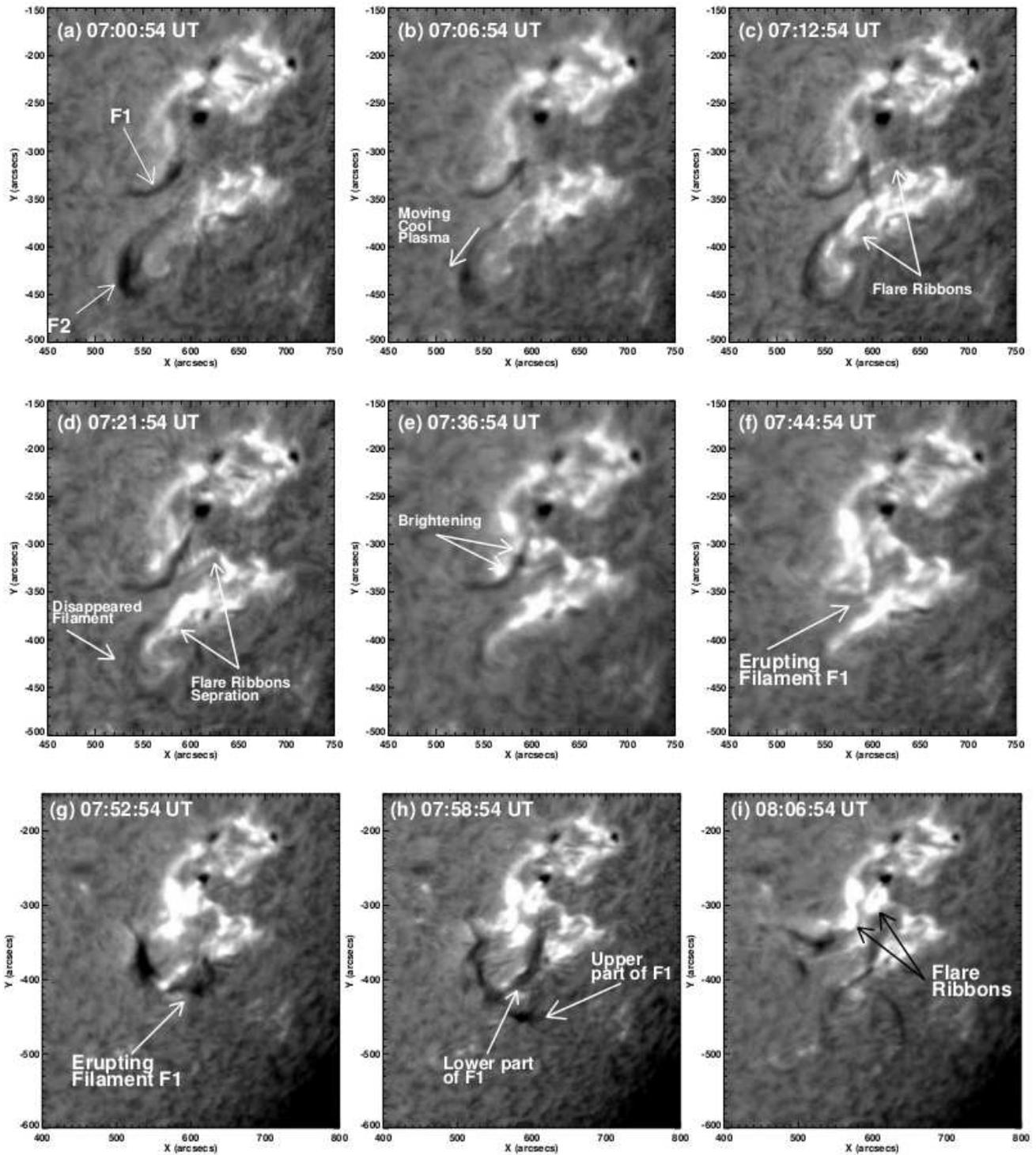}
	}
\vspace*{-2cm}
\caption{Sequence of selected NSO/GONG H$\alpha$ images showing the overall evolution of the filaments activities.}
\label{fig4}
\end{figure}

\begin{figure}
\vspace*{-4cm}
\centerline{
	\hspace*{0.0\textwidth}
	\includegraphics[width=2.2\textwidth,clip=]{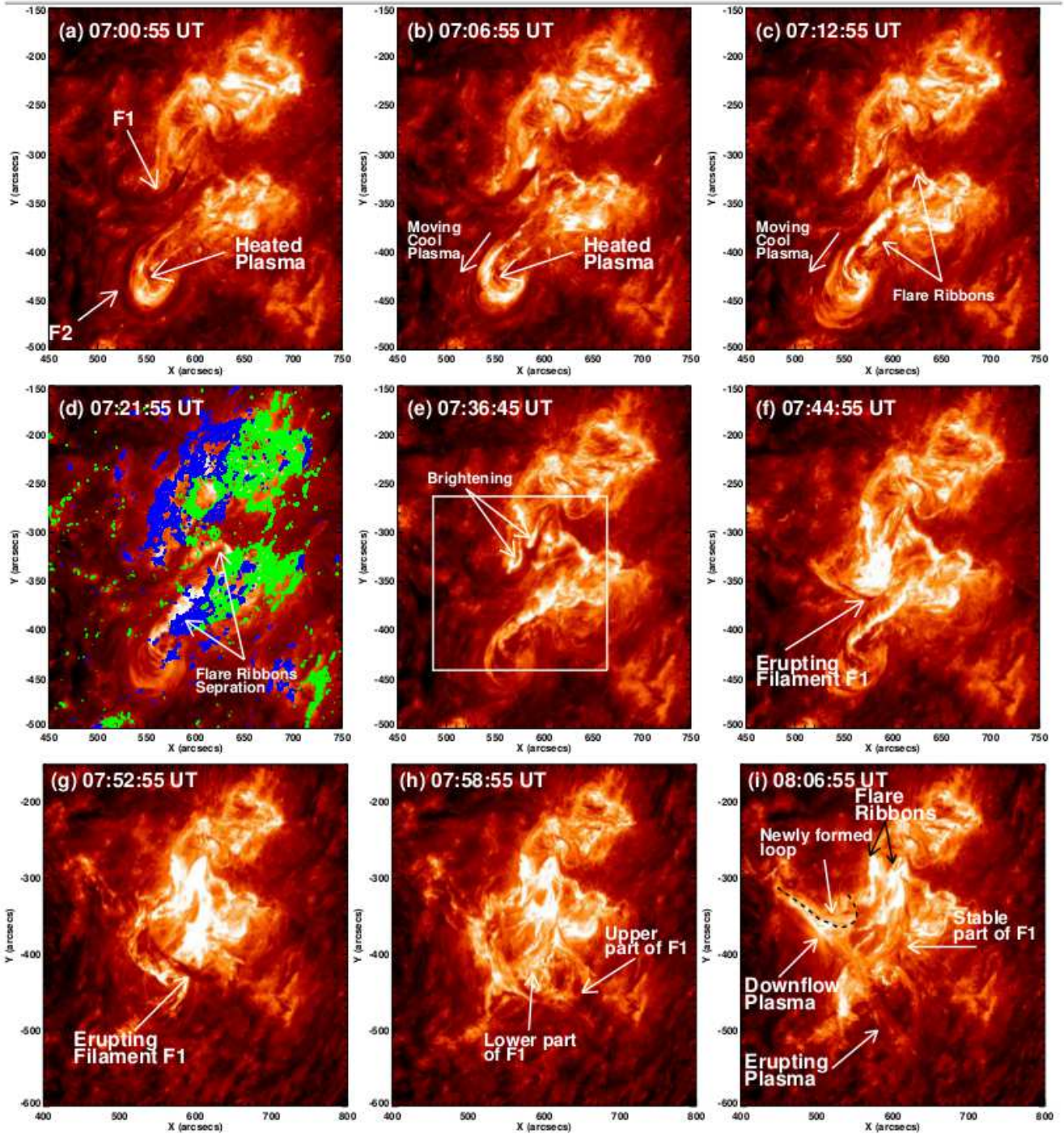}
	}
\vspace*{-2cm}
\caption{Sequence of selected \textit{SDO}/AIA 304 \AA~images showing the overall evolution of the filaments activities. The white box in panel (e) shows the area used for intensity profiles. The newly formed loop system are shown by the black dotted lines in panel (i). The green and blue contours in panel (d) are the \textit{SDO}/HMI magnetic field contours for positive and negative polarity, respectively. The contour levels are $\pm50,\pm100,\pm200,\pm300,\pm400,\pm500$ Gauss.}
\label{fig5}
\end{figure}

\begin{figure}
\vspace*{-4cm}
\centerline{
	\hspace*{0.0\textwidth}
	\includegraphics[width=2.2\textwidth,clip=]{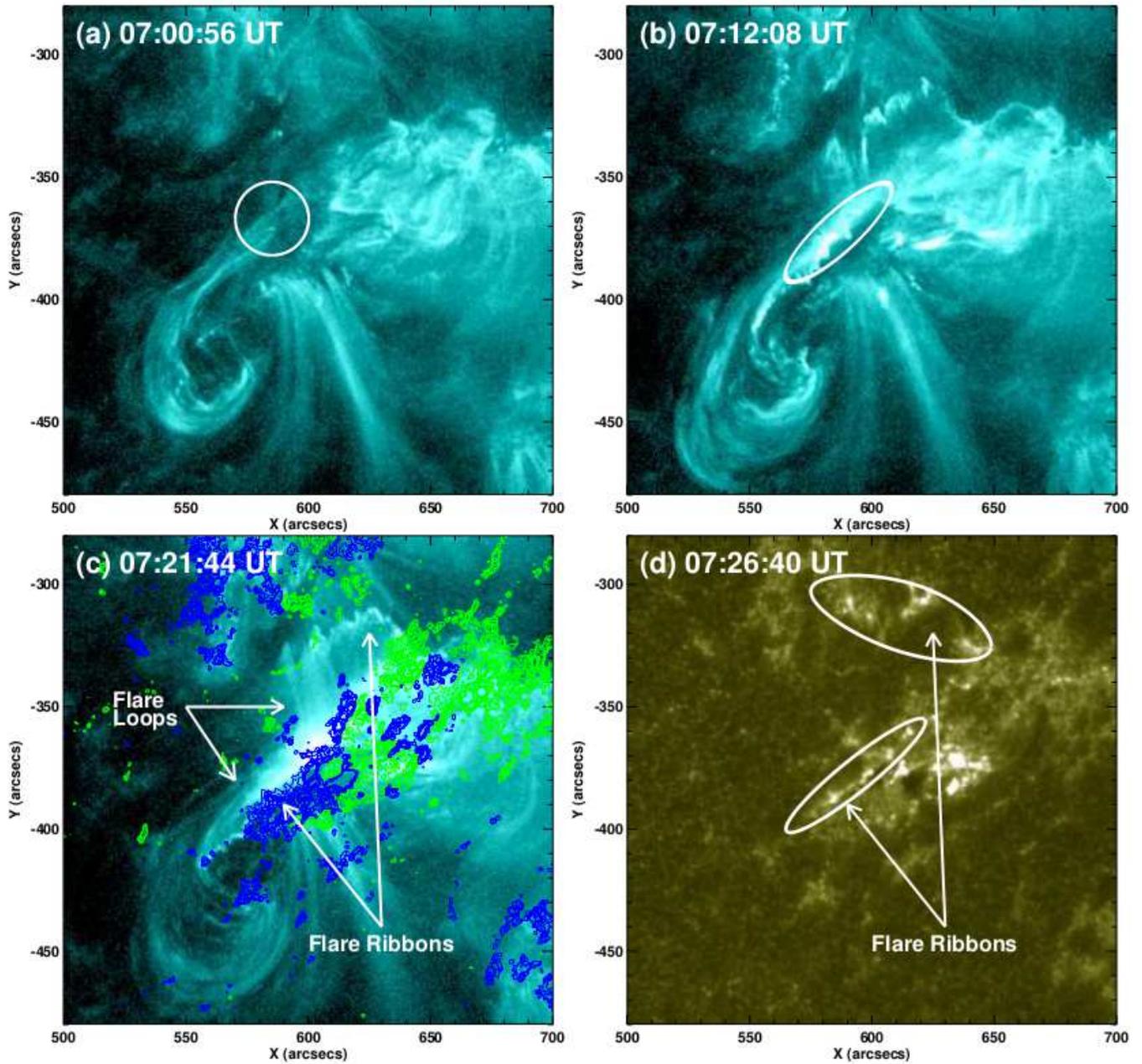}
	}
\vspace*{-2.5cm}
\caption{(a)--(c) \textit{SDO}/AIA 131 \AA~images showing the occurrence of the first flare. The possible region of reconnection is marked with circle and ellipse in panels (a) and (b) respectively. (d) \textit{SDO}/AIA 1600 \AA~images showing the ribbons of the first flare, marked by the white ellipses. The green and blue contours in panel (c) are the \textit{SDO}/HMI magnetic field contours for positive and negative polarity, respectively. The contour levels are $\pm50,\pm100,\pm200,\pm300,\pm400,\pm500$ Gauss.}
\label{fig6}
\end{figure}

\begin{figure}
\vspace*{-2cm}
\centerline{
	\hspace*{0.0\textwidth}
	\includegraphics[width=1.7\textwidth,clip=]{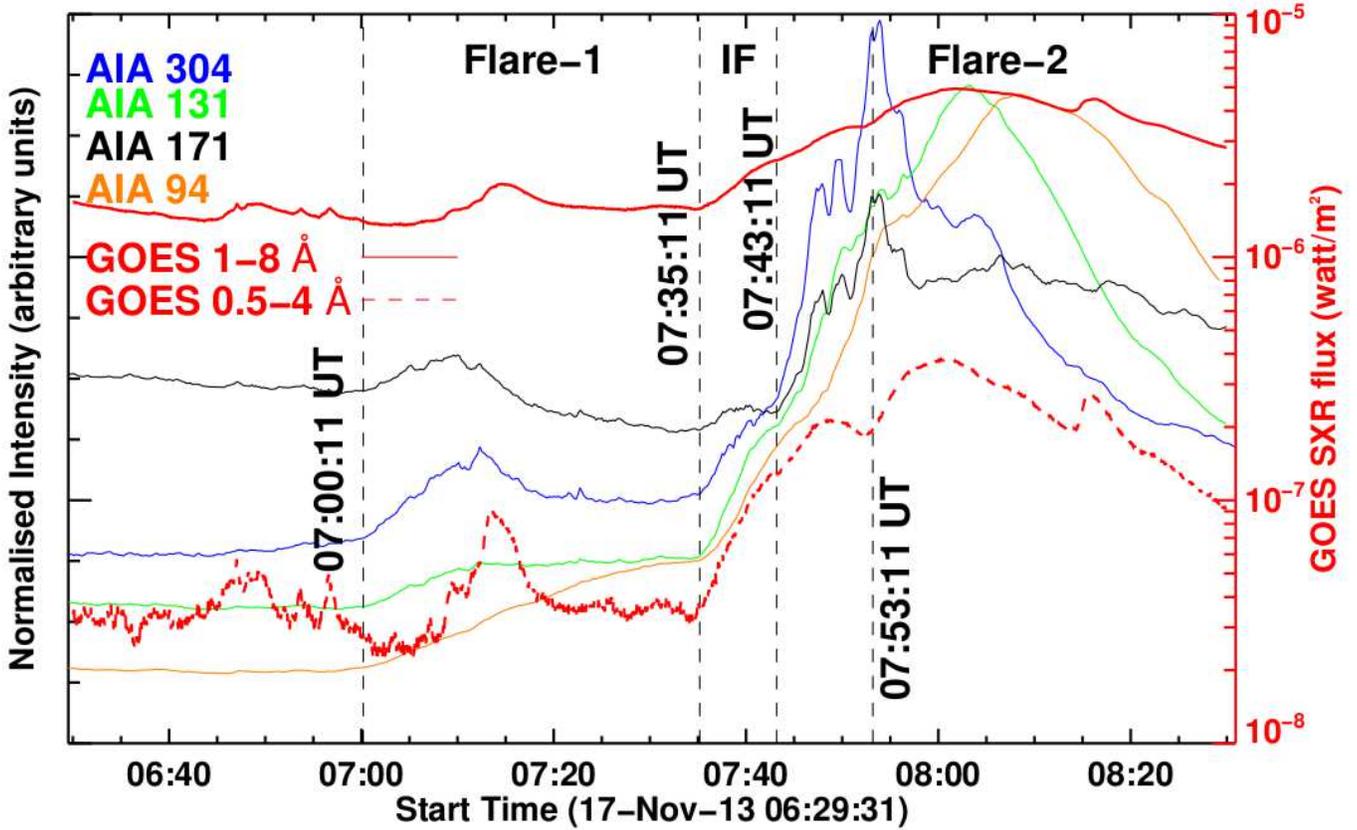}
	}
\vspace*{-2.8cm}
\caption{Intensity profiles of the flaring region in \textit{SDO}/AIA 304 (blue), 171 (black), 131 (green) and 94 (orange) \AA~wavelength channels. The overplotted solid red and dashed red lines represent the \textit{GOES} 1--8 and 0.4--5 \AA~profiles for the comparison. Different flare phases are also marked. IF represents the intermediate phase between the first and second flares. The vertical dashed lines from left to rignt correspond to the 07:00 UT, 07:35 UT and 07:43 UT, respectivley.}
\label{fig7}
\end{figure}

\begin{figure}
\vspace*{-2cm}
\centerline{
	\hspace*{0.0\textwidth}
	\includegraphics[width=2\textwidth,clip=]{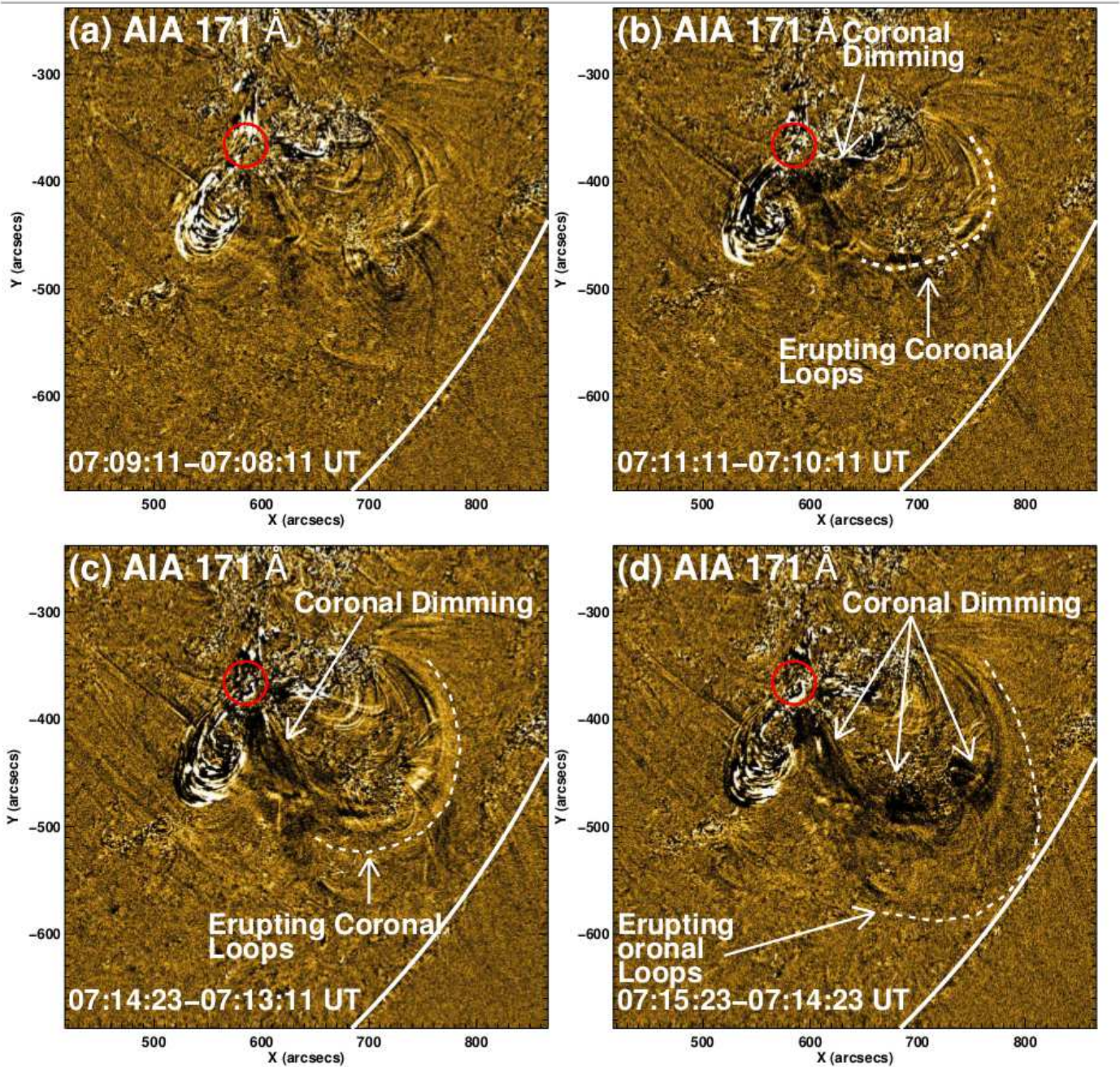}
	}
\vspace*{-1.6cm}
\caption{\textit{SDO}/AIA 171 \AA~running difference images showing the eruption of overlying field lines during the first flare. The front of the eruption is represented by the dashed curved lines. White curved line represents the limb of the Sun. The reconnection regions are shown by the red circles.}
\label{fig8}
\end{figure}

\begin{figure}
\vspace*{-4.5cm}
\centerline{
	\hspace*{0.0\textwidth}
	\includegraphics[width=2.6\textwidth,clip=]{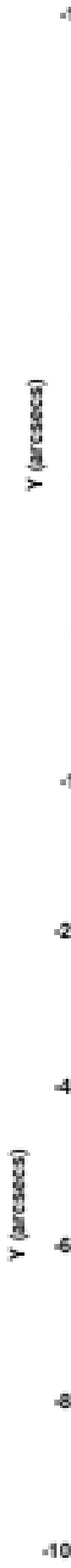}
	}
\vspace*{-2.8cm}
\caption{(a)--(e) \textit{STEREO-A}/EUVI 195 \AA~running difference images during 07:05:30 UT to 07:40:30 UT, showing the eruption of overlying coronal loops during the first flare. White curved line represents the limb of the Sun. (f) Height--time profile of the erupting loops.}
\label{fig9}
\end{figure}

\begin{figure}
\vspace*{-4cm}
\centerline{
	\hspace*{0.0\textwidth}
	\includegraphics[width=2.2\textwidth,clip=]{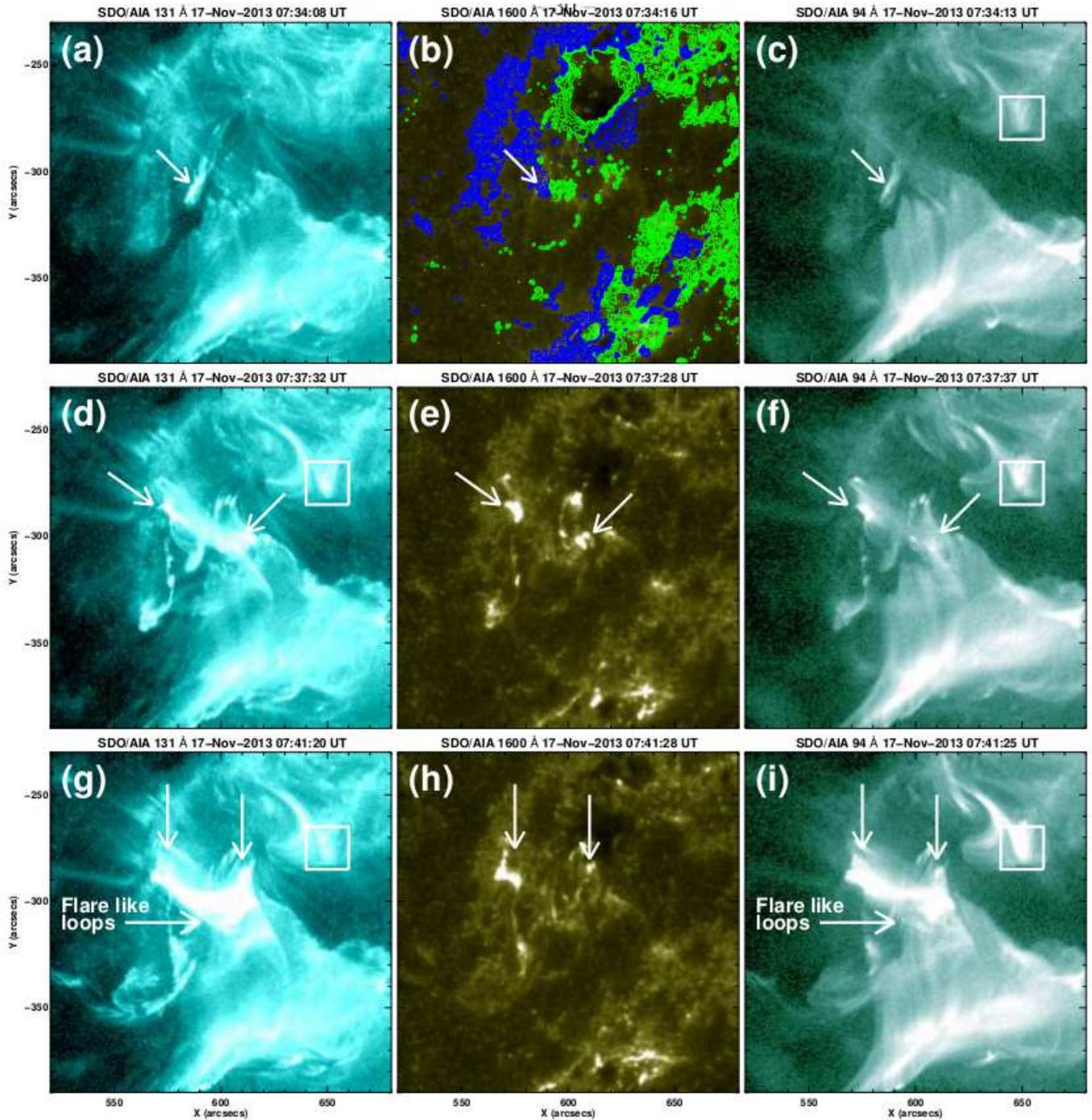}
	}
\vspace*{-2cm}
\caption{\textit{SDO}/AIA 131 (left column), 1600 (middle column) and 94 (right column) \AA\ images showing the evidence of second stage reconnection. The upper, middle and lower panels show the images at $\sim$07:34 UT, $\sim$07:37 UT and $\sim$07:41 UT, respectively. The box in the top right show the area near box 2 in Figure~\ref{fig3}, where the field lines of both the system lie close to each other. The green and blue contours in panel (b) are the \textit{SDO}/HMI magnetic field contours for positive and negative polarity, respectively. The contour levels are $\pm50,\pm100,\pm200,\pm300,\pm400,\pm500$ Gauss.}
\label{fig10}
\end{figure}

\begin{figure}
\vspace*{-5cm}
\centerline{
	\hspace*{0.0\textwidth}
	\includegraphics[width=2.5\textwidth,clip=]{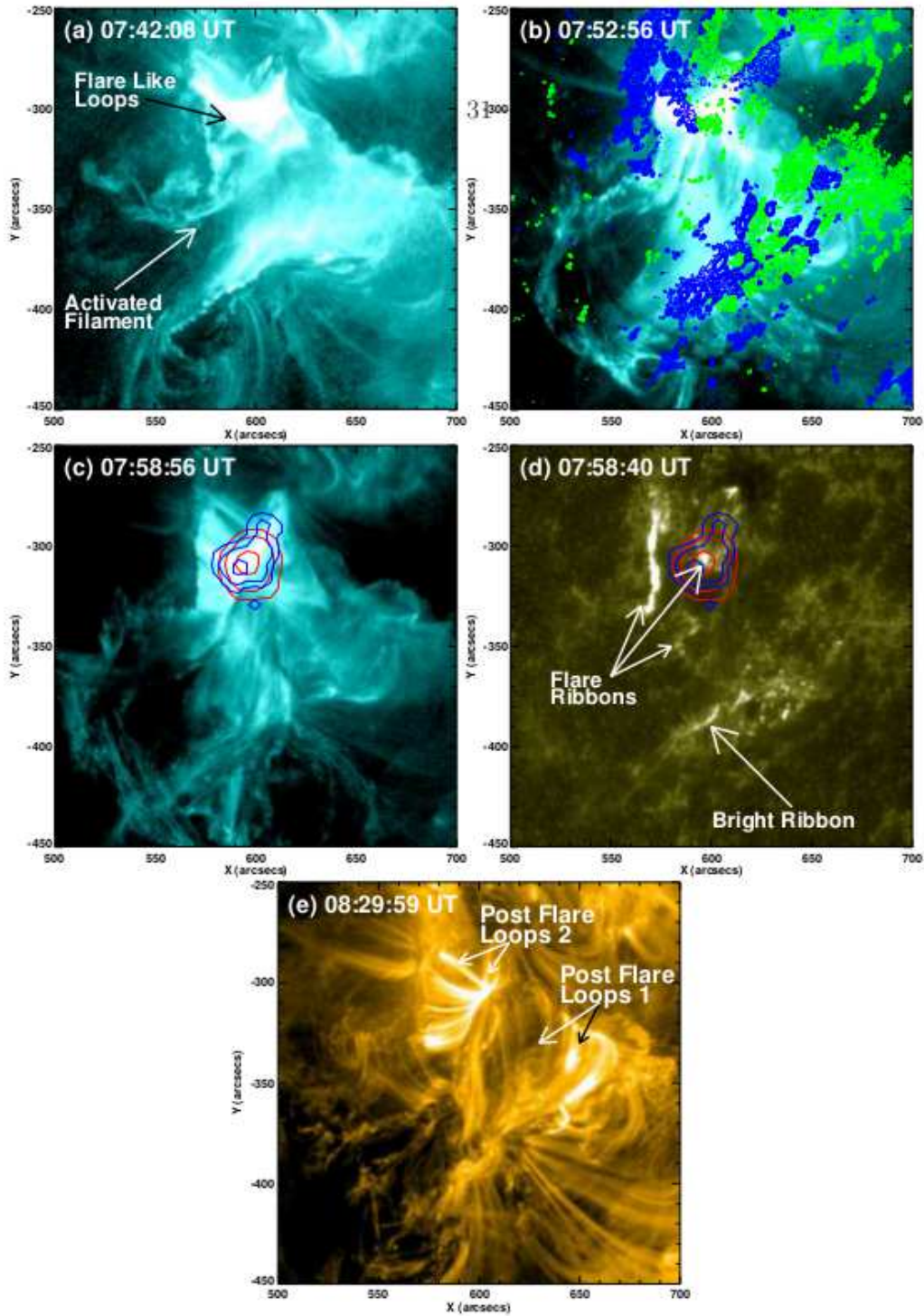}
	}
\vspace*{-3cm}
\caption{(a)--(c) \textit{SDO}/AIA 131 \AA~images showing the evolution of the filament eruption and the triggerting of second flare at 07:42:08 UT, 07:52:56 UT and 07:58:56 UT, respectively. The green and blue contours in panel (b) are the \textit{SDO}/HMI magnetic field contours for positive and negative polarity, respectively. (d) \textit{SDO}/AIA 1600 \AA~image at 07:58:40 UT showing the flare ribbons of the second flare. The contour levels are $\pm50,\pm100,\pm200,\pm300,\pm400,\pm500$ Gauss. The contours in panels (c) and (d) are the \textit{RHESSI} 6--12 KeV (red) and 12--25 KeV (blue) contours showing the location of the flare. The contour levels are 70\%, 80\% and 90\% and the integration time is 60 seconds. (e) \textit{SDO}/AIA 171 \AA~image at 08:29:59 UT showing the post flare loops associated with the two flares.}
\label{fig11}
\end{figure}

\begin{figure}
\vspace*{-4cm}
\centerline{
	\hspace*{0.0\textwidth}
	\includegraphics[width=2.3\textwidth,clip=]{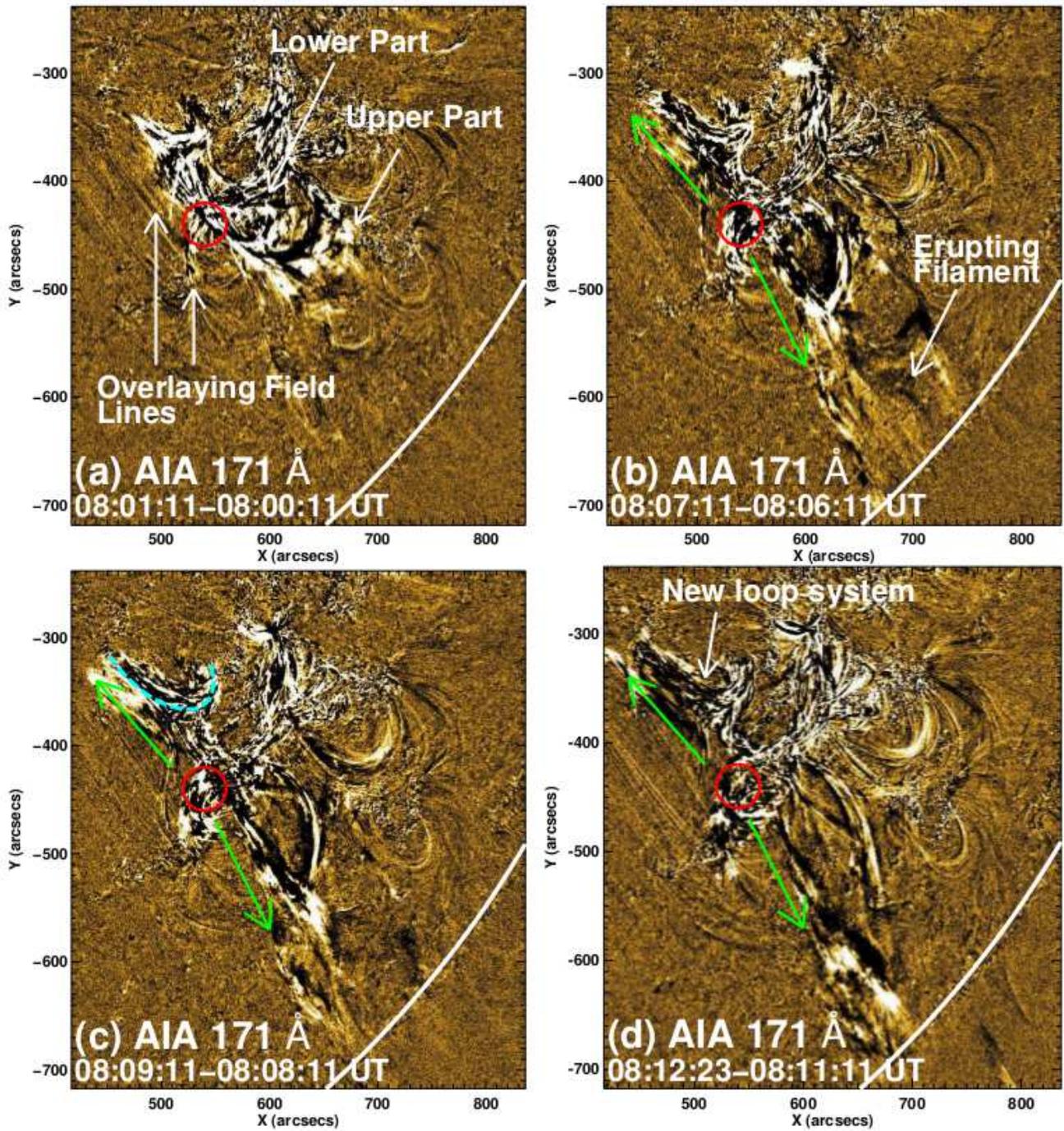}
	}
\vspace*{-2.5cm}
\caption{\textit{SDO}/AIA 171 \AA~running difference images during 08:01:11 UT to 08:12:23 UT showing the eruption of filament F1 and its reconnection with the surrounding ambient medium. The green arrows represent the upward and downward plasma flows from the reconnection region. The reconnection regions are shown by the red circles. White curved line represents the limb of the Sun. The new formed loop system is represented by the cyan color loops in panels (c) and (d).}
\label{fig12}
\end{figure}

\begin{figure}
\vspace*{-6cm}
\centerline{
	\hspace*{0.0\textwidth}
	\includegraphics[width=2.8\textwidth,clip=]{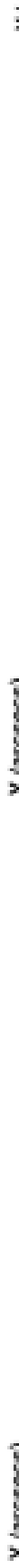}
	}
\vspace*{-3.8cm}
\caption{Left panel (a)--(c): \textit{SDO}/AIA 304 \AA~running difference images during  07:51 UT to 08:18 UT showing the on disk observations of the eruption of filament F1 eruption. Right panel (d)--(f): \textit{STEREO-A}/EUVI 304 \AA~images during 07:51:15 UT to 08:18:45 UT represent the limb observations of the eruption of filament F1. Arrows mark the leading edge of the erupting part. Dashed line in panels (c) and (f) show the rough trajectory along which the eruption of filament F1 has been tracked for the height measurement. White curved lines represent the limb of the Sun.}
\label{fig13}
\end{figure}

\begin{figure}
\vspace*{-2cm}
\centerline{
	\hspace*{0.0\textwidth}
	\includegraphics[width=1.5\textwidth,clip=]{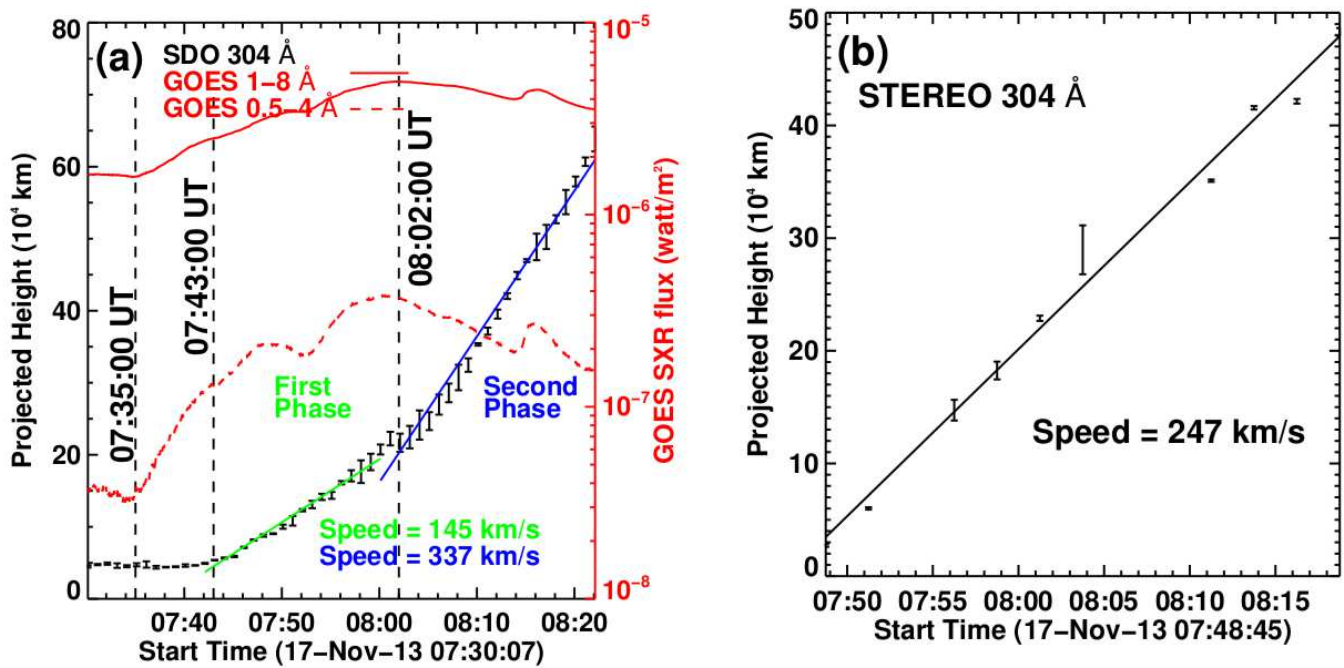}
	}
\vspace*{-2.5cm}
\caption{Projected height--time profile of the filament F1 eruption derived by \textit{SDO}/AIA 304 \AA~(a) and \textit{STEREO-A}/EUVI 304 \AA~(b) images. Red solid and dashed curves in panel (a) show the \textit{GOES} X-ray profiles in 1--8 and 0.5--4 \AA~wavelength channels. The first vertical dashed line in panel (a) shows the time of compact brightening under the filament F1 at around 07:35 UT. The second vertical dashed line represents the start of filament F1 eruption at around 07:43 UT. The third line represents the second stage fast eruption of filament F1 eruption at around 08:02 UT. Rough trajectory along which the height measurement is performed is shown in Figures~\ref{fig13}(c) and~\ref{fig13}(f) by the dashed white line. The error bars are the standard deviations estimated using three repeated measurements of the same point.}
\label{fig14}
\end{figure}

Another interesting result is that filament F1 broke into two different parts, i.e., the erupting upper and stable lower parts (Figures~\ref{fig4}(h),~\ref{fig5}(h) and~\ref{fig12}(a)). We suggest that the reconnection of the erupting filament F1 with the surviving field of F2 may break F1 into two parts. Also, the observed enhancement in the brightness of the southern ribbon of the first flare during the eruption of filament F1 also provides some hint for this type of interaction (Figure~\ref{fig11}(d)).

\begin{figure}
\vspace*{-2cm}
\centerline{
	\hspace*{0.0\textwidth}
	\includegraphics[width=1.5\textwidth,clip=]{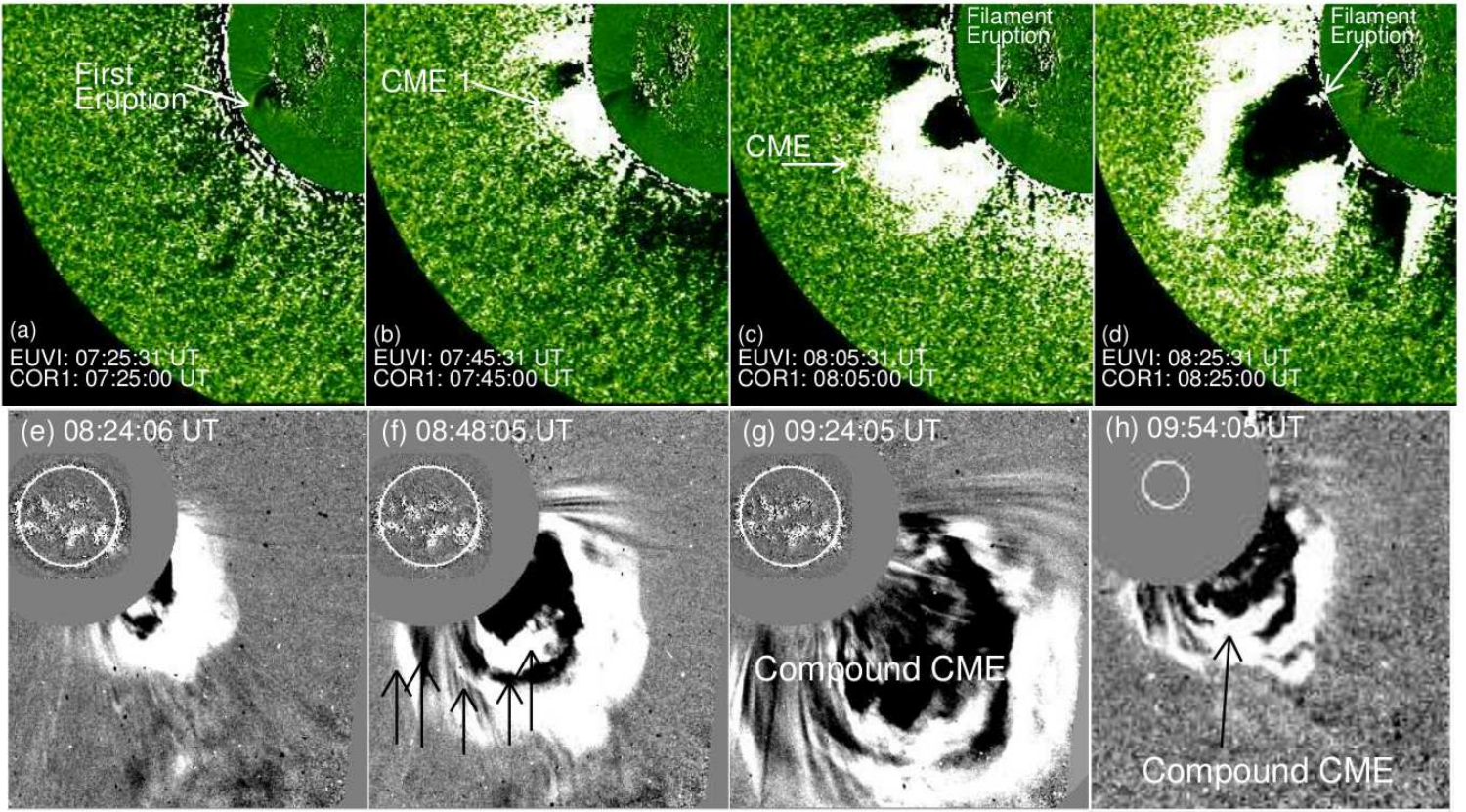}
	}
\vspace*{-2cm}
\caption{\textit{STEREO-A}/COR1 ((a)--(d)) and \textit{SOHO}/LASCO C2 and C3 ((e)--(f)) images showing the sequence of CMEs associated with first and second eruptions during 07:25 UT to 09:54 UT on 2013 November 17. The difference images are collected from CDAW stereo catalog (\url{http://cdaw.gsfc.nasa.gov/stereo/daily\_movies/2013/11/17/index.html}) as well as from \textit{SOHO}/LASCO catalog (\url{http://cdaw.gsfc.nasa.gov/CME\_list/}). The black arrows in panel (f) represent the multiple white and black bands, which show the evidence of different parts of two CMEs.}
\label{fig15}
\end{figure}

\begin{figure}
\vspace*{-2cm}
\centerline{
	\hspace*{0.0\textwidth}
	\includegraphics[width=2\textwidth,clip=]{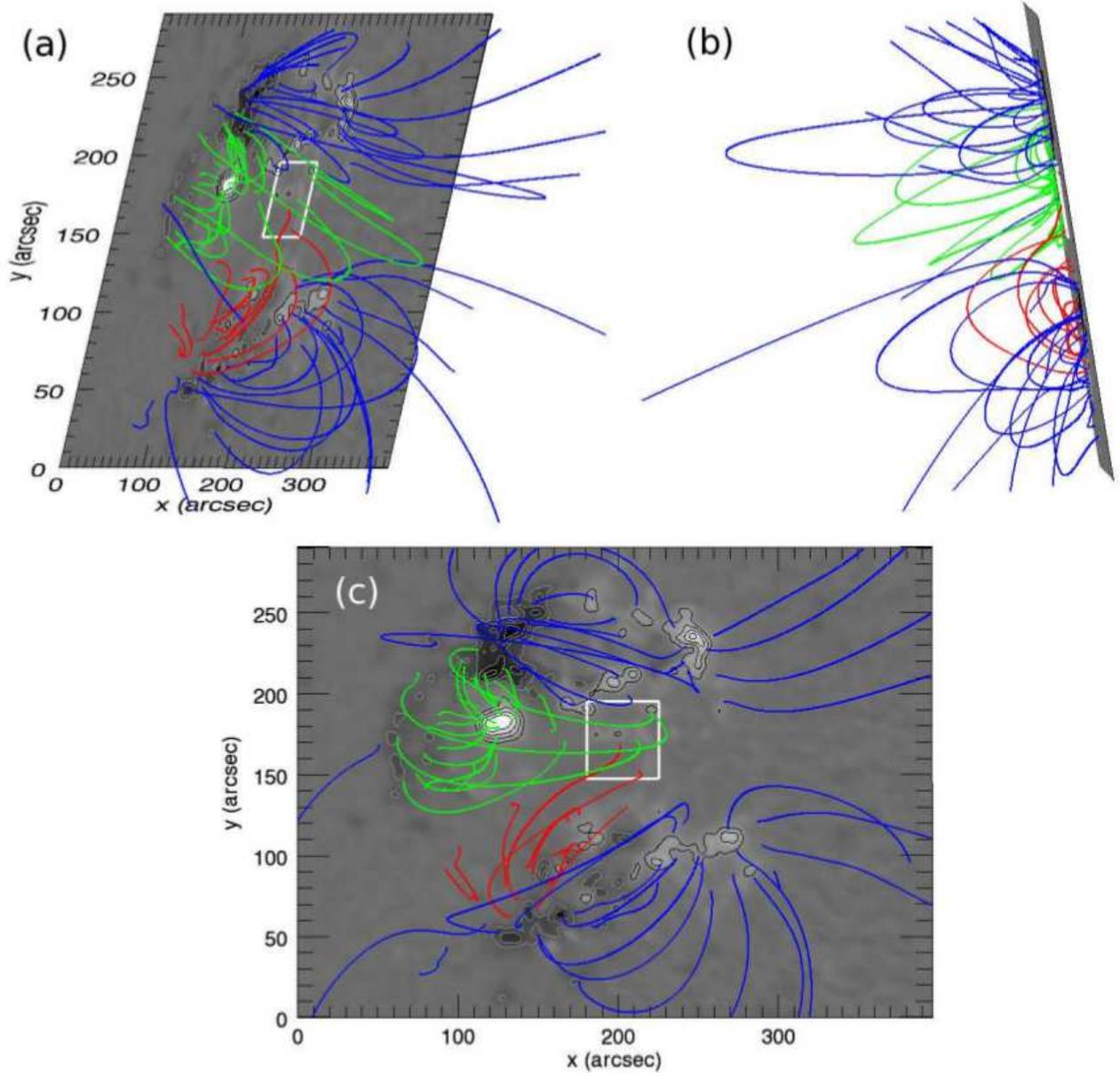}
	}
\vspace*{-2.5cm}
\caption{NLFFF extrapolation of the two active region photospheric HMI magnetic field at 07:00:00 UT on 2013 November 17 viewed in three different angles, i.e., (a) the AIA view, (b) approximate \textit{STEREO} view, and (c) top view. The white box is  the approximate box 2 in Figure~\ref{fig3}. It shows  the positive polarities where the overlying arcades of F1 (green lines) and those of F2 (red lines) are anchored.}
\label{fig16}
\end{figure}

\begin{figure}
\vspace*{-4cm}
\centerline{
	\hspace*{0.0\textwidth}
	\includegraphics[width=2.5\textwidth,clip=]{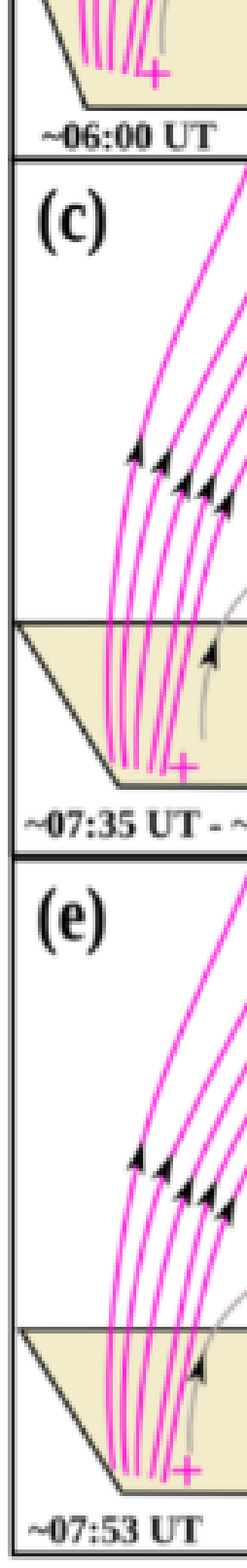}
	}
\vspace*{-2.5cm}
\caption{Schematic representation of the sympathetic eruptions and associated flares. Blue and pink lines represent the nearby ambient magnetic field lines. Green and red lines show the overlying arcades of filaments F1 and F2 respectively. Black lines represent the reconnected field lines. Brown color lines show the flare ribbons. The approximate times corresponding to the real observations are mentioned in each panel.}
\label{fig17}
\end{figure}

We also observe that the second fast eruption (speed $\sim$377 $\rm km~s^{-1}$) overtook the first slow eruption (160 $\rm km~s^{-1}$) and the associated CMEs merged and formed a compound CME (see Section~\ref{sect37} and Figure~\ref{fig15}). The compound CME structure can be seen at 08:48 UT and onward with multiple leading edges and cavities (Figures~\ref{fig15}(f)--(h)). These observations suggest that the sympathetic nearby eruptions and associated CMEs in the same directions can also lead to the interesting CME--CME interaction events in the corona near the Sun. These interactions have great importance for the production of solar energetic particle events \citep[e.g.,][]{Gopalswmay01,Joshi13b,Masson13}.

Study of nearby eruptions, flares and CMEs tells important information about magnetic interactions and reconnections between them. We have identified several physical mechanisms for producing this sequence of sympathetic events in two different active regions: canceling flux, shear arcades, inflation or expansion of magnetic field systems. In this study, all these mechanisms work together or successively between the different magnetic systems. It provides a good example of sympathetic events. These kind of observational studies also provide important inputs for MHD modelings of sympathetic eruptions and the associated flares. High resolution observations and more numerical simulation studies are needed to understand the sympathetic events more clearly.

\acknowledgments
The authors thank the referee for providing valuable comments and suggestions. We thank \textit{SDO}/AIA, \textit{SDO}/HMI, \textit{STEREO}/SECCHI, NSO/GONG, \textit{SOHO}/LASCO and \textit{RHESSI} teams for providing their data for the present study. This work is supported by the BK21 plus program through the National Research Foundation (NRF) funded by the Ministry of Education of Korea. NCJ thank School of Space Research, Kyung Hee University for providing the postdoctoral grant. YG is supported by NSFC 11203014 and BELSPO postdoctoral grant. YG thanks the Observatoire de Paris. We are thankful to Dr. Tibor T{\"o}r{\"o}k for their valuable suggestions and comments on the present work.


\clearpage

\clearpage

\clearpage

\clearpage

\clearpage

\clearpage

\clearpage

\clearpage

\clearpage


\clearpage


\clearpage

\clearpage

\clearpage

\clearpage

\clearpage

\clearpage

\clearpage

\end {document}